\documentclass[fleqn,twoside]{article}
\usepackage{espcrc1}

\newcommand{\AmS}{{\protect\the\textfont2
  A\kern-.1667em\lower.5ex\hbox{M}\kern-.125emS}}

\title{On balanced clustering with tree-like structures over clusters}

\author{Mark Sh. Levin
\thanks{
 Mark Sh. Levin:~
 Inst. for Inform. Transmission Problems,
 Russian Academy of Sciences;
  http://www.mslevin.iitp.ru;
 email: mslevin@acm.org
  } }

\begin{document}

\maketitle

\begin{abstract}
 The article addresses balanced clustering
 problems with an additional requirement as
 a tree-like structure over the obtained balanced clusters.
 This kind of clustering problems
 can be useful in some applications
 (e.g., network design, management and routing).
 Various types of the initial elements are considered.
 Four basic greedy-like solving strategies (design framework) are considered:
 balancing-spanning strategy,
 spanning-balancing strategy,
 direct strategy, and
 design of layered structures with balancing.
 An extended description of
 the spanning-balancing strategy is
 presented including four solving schemes and
 an illustrative numerical example.

~~


\vspace{1pc}
\end{abstract}

\maketitle

\tableofcontents

\newcounter{cms}
\setlength{\unitlength}{1mm}

%
\section{Introduction}

 Balancing clustering is often a significant part of many
 theoretical and practical problems.
 In this paper, special balanced clustering
 problems with an additional requirement
 as
 a tree-like structure over the obtained balanced clusters
 are considered.
 This kind of clustering problems
 can be useful in many network applications
 (e.g., network design, management and routing).
 The global quality of the solution is examined as quality of the obtained balanced
 clustering solution
 and quality of the spanning structure over the balanced
 clustering solution (i.e., two-component vector).
 Thus, the resultant problem is targeted to find the Pareto-efficient
 solution(s).
  Balanced clustering problems are often examined for the following basic cases
  (e.g., \cite{lev17bal1,lev17balj}):
 (i) cluster size  requirement,
 (ii) cluster weight balance requirement, and
  (iii) cluster element structure requirement
  (here several types of the initial elements are considered).
 In the paper, the third case above is mainly under examination.
 Here, the solution quality is based on multiset estimate
  \cite{lev12a,lev15,lev17bal1,lev17balj}.
 The following kinds of  basic spanning trees
 can be considered:
 degree balanced tree and
 height balanced tree.
 Quality of the balanced structure is a proximity between
 obtained spanning structure and the basic type of balanced
 structure.
 The considered problems are illustrated by numerical examples.

 Four basic greedy-like solving strategies are considered:
 balancing-spanning strategy,
 spanning-balancing strategy,
 direct strategy, and
 design of layered structures with balancing.
 An extended description of
 the spanning-balancing strategy is
 presented including four solving schemes and
 an illustrative numerical example.
 Approaches for improvement of the obtained solutions
 are briefly pointed out.

\section{Problem description and statement}

  A general glance to balanced clustering problems
  has been presented in
 \cite{lev17bal1,lev17balj}.
 Here a special balanced clustering problem is examined when
 the solution involves balanced clustering solution and a
 spanning structure (e.g., tree) over the obtained balanced
 clusters (Fig. 1).

\begin{center}
\begin{picture}(115,38.5)
\put(012,00){\makebox(0,0)[bl]{Fig. 1.
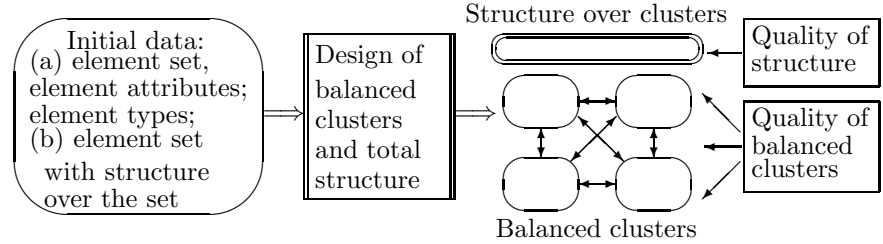
 Balanced clustering with structure over clusters}}
\put(16.5,22){\oval(33,26)}

\put(07,31){\makebox(0,0)[bl]{Initial data:}}
\put(02,27.5){\makebox(0,0)[bl]{(a) element set, }}
\put(02,24){\makebox(0,0)[bl]{element attributes;}}
\put(02,20.5){\makebox(0,0)[bl]{element types;}}
\put(02,17){\makebox(0,0)[bl]{(b) element set  }}
\put(04,13.5){\makebox(0,0)[bl]{with structure }}
\put(04,10){\makebox(0,0)[bl]{over the set }}

\put(33,21.5){\makebox(0,0)[bl]{\(\Longrightarrow\)}}

\put(38.5,11){\line(1,0){20}} \put(38.5,33){\line(1,0){20}}
\put(38.5,11){\line(0,1){22}} \put(58.5,11){\line(0,1){22}}
\put(39,11){\line(0,1){22}} \put(58,11){\line(0,1){22}}

\put(40,28.5){\makebox(0,0)[bl]{Design of }}
\put(40,24.5){\makebox(0,0)[bl]{balanced }}
\put(40,20.5){\makebox(0,0)[bl]{clusters }}
\put(40,16.5){\makebox(0,0)[bl]{and total}}
\put(40,12.5){\makebox(0,0)[bl]{structure}}

\put(58.5,21.5){\makebox(0,0)[bl]{\(\Longrightarrow\)}}

\put(60,34.5){\makebox(0,0)[bl]{Structure over clusters}}

\put(77.5,31){\oval(28,04)} \put(77.5,31){\oval(27,03)}


\put(70,24){\oval(10,07)} \put(85,24){\oval(10,07)}

\put(75,24){\vector(1,0){05}} \put(80,24){\vector(-1,0){05}}

\put(70,13){\oval(10,07)} \put(85,13){\oval(10,07)}

\put(75,13){\vector(1,0){05}} \put(80,13){\vector(-1,0){05}}

\put(70,16.5){\vector(0,1){04}} \put(70,20.5){\vector(0,-1){04}}
\put(85,16.5){\vector(0,1){04}} \put(85,20.5){\vector(0,-1){04}}

\put(74,16){\vector(1,1){06}} \put(80,22){\vector(-1,-1){06}}
\put(75,22){\vector(1,-1){06}} \put(81,16){\vector(-1,1){06}}

\put(64,06){\makebox(0,0)[bl]{Balanced clusters}}

\put(98,31){\makebox(0,0)[bl]{Quality of}}
\put(98,28){\makebox(0,0)[bl]{structure}}

\put(97,26){\line(1,0){18}} \put(97,35){\line(1,0){18}}
\put(97,26){\line(0,1){09}} \put(115,26){\line(0,1){09}}

\put(96.7,30.5){\vector(-1,0){04.5}}

\put(98,20){\makebox(0,0)[bl]{Quality of}}
\put(98,17){\makebox(0,0)[bl]{balanced }}
\put(98,14){\makebox(0,0)[bl]{clusters }}

\put(97,12){\line(1,0){18}} \put(97,24){\line(1,0){18}}
\put(97,12){\line(0,1){12}} \put(115,12){\line(0,1){12}}

\put(96.5,20){\vector(-1,1){05}} \put(96.7,18){\vector(-1,0){05}}
\put(96.5,16){\vector(-1,-1){05}}

\end{picture}
\end{center}

  Some designations are as follows
 (e.g., \cite{lev12a,lev15,lev15c,lev15d,lev17bal1,lev17balj}).

 Let \(A=\{a_{1},...,a_{i},...,a_{n}\}\) be an initial set of
 items;
 \(\widetilde{X} = \{X_{1},...,X_{j},...X_{k}\}\)
  be a clustering solution where \(X_{j} \subseteq A\)
 (for simplification without intersections:  \( |X_{j_{1}} \cap X_{j_{1}}|=0
  ~~ \forall ~ 1\leq j_{1} < j_{2}  \leq k \)).
 The following types of items are considered:~
 type \(1\),..., type \(\xi\), ...,type \(l\).
 The description of  cluster \(X_{j}\) by its element types is
 (vector, multiset estimate
 \cite{lev12a,lev15,lev17bal1,lev17balj}):~~
%
%
%
 \(
 e (X_{j}) = ( \eta_{1}(X_{j}),...,\eta_{\xi}(X_{j}),...,\eta_{l}(X_{j}) )
  \)
 where \(\eta_{\xi}(X_{j})\) is the number of elements of type \( \xi\)
 which are contained in cluster \(X_{j}\),
  \(\sum_{\xi = \overline{1,l}}~\eta_{\xi}(X_{j})=|X_{j}|\).

 Fig. 2 depicts four types of items (elements) and several
  cluster examples.
 Here, descriptions of the clusters by element types are:
 \(e(X_{1}) = (0,0,0,4 ) \),
 \(e(X_{2}) = (0,0,0,3 ) \),
 \(e(X_{3}) = (0,0,1,3 ) \),
 \(e(X_{4}) = (0,0,1,2 ) \),
 \(e(X_{5}) = (0,1,2,1 ) \),
 \(e(X_{6}) = (0,1,1,1 ) \),
 \(e(X_{7}) = (1,1,1,1 ) \), and
 \(e(X_{8}) = (0,1,2,0 ) \).

\begin{center}
\begin{picture}(115,35)
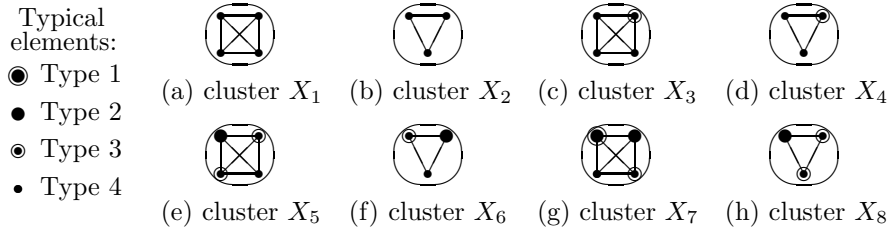
\put(010,00){\makebox(0,0)[bl]{Fig. 2. Illustrative example:
 typical elements and clusters }}


\put(01,31){\makebox(0,0)[bl]{Typical }}
\put(00,28.5){\makebox(0,0)[bl]{elements: }}

\put(01,25){\circle*{1.6}} \put(01,25){\circle{2.6}}
\put(04,23.5){\makebox(0,0)[bl]{Type \(1\) }}

\put(01,20){\circle*{1.0}} \put(01,20){\circle*{1.8}}
\put(04,18.5){\makebox(0,0)[bl]{Type \(2\) }}

\put(01,15){\circle*{1.0}} \put(01,15){\circle{1.8}}
\put(04,13.5){\makebox(0,0)[bl]{Type \(3\) }}

\put(01,10){\circle*{1.0}} \put(01,10){\circle*{1.0}}
\put(04,08.5){\makebox(0,0)[bl]{Type \(4\) }}

\put(30.5,30.5){\oval(09,08)}

\put(28,33){\circle*{1.0}} \put(33,33){\circle*{1.0}}
\put(28,28){\circle*{1.0}}\put(33,28){\circle*{1.0}}

\put(28,28){\line(1,0){05}} \put(28,33){\line(1,0){05}}
\put(28,28){\line(0,1){05}} \put(33,28){\line(0,1){05}}

\put(28,28){\line(1,1){05}} \put(33,28){\line(-1,1){05}}

\put(20,21){\makebox(0,0)[bl]{(a) cluster \(X_{1}\) }}

\put(55.5,30.5){\oval(09,08)}

\put(53,33){\circle*{1.0}} \put(58,33){\circle*{1.0}}

\put(55.5,28){\circle*{1.0}}

\put(53,33){\line(1,0){05}}

\put(55.5,28){\line(-1,2){02.5}} \put(55.5,28){\line(1,2){02.5}}

\put(45,21){\makebox(0,0)[bl]{(b) cluster \(X_{2}\) }}

\put(80.5,30.5){\oval(09,08)}

\put(78,33){\circle*{1.0}}

\put(83,33){\circle*{1.0}} \put(83,33){\circle{1.8}}
\put(78,28){\circle*{1.0}}\put(83,28){\circle*{1.0}}

\put(78,28){\line(1,0){05}} \put(78,33){\line(1,0){05}}
\put(78,28){\line(0,1){05}} \put(83,28){\line(0,1){05}}

\put(78,28){\line(1,1){05}} \put(83,28){\line(-1,1){05}}

\put(70,21){\makebox(0,0)[bl]{(c) cluster \(X_{3}\) }}

\put(105.5,30.5){\oval(09,08)}

\put(103,33){\circle*{1.0}}

\put(108,33){\circle*{1.0}} \put(108,33){\circle{1.8}}
\put(105.5,28){\circle*{1.0}}

\put(103,33){\line(1,0){05}}

\put(105.5,28){\line(-1,2){02.5}} \put(105.5,28){\line(1,2){02.5}}

\put(95,21){\makebox(0,0)[bl]{(d) cluster \(X_{4}\) }}

\put(30.5,14.5){\oval(09,08)}

\put(28,17){\circle*{1.0}} \put(28,17){\circle*{1.8}}

\put(33,17){\circle*{1.0}} \put(33,17){\circle{1.8}}

\put(28,12){\circle*{1.0}} \put(28,12){\circle{1.8}}

\put(33,12){\circle*{1.0}}

\put(28,12){\line(1,0){05}} \put(28,17){\line(1,0){05}}
\put(28,12){\line(0,1){05}} \put(33,12){\line(0,1){05}}

\put(28,12){\line(1,1){05}} \put(33,12){\line(-1,1){05}}

\put(20,05){\makebox(0,0)[bl]{(e) cluster \(X_{5}\) }}

\put(55.5,14.5){\oval(09,08)}

\put(53,17){\circle*{1.0}} \put(53,17){\circle{1.8}}
\put(58,17){\circle*{1.0}} \put(58,17){\circle*{1.8}}

\put(55.5,12){\circle*{1.0}}

\put(53,17){\line(1,0){05}}

\put(55.5,12){\line(-1,2){02.5}} \put(55.5,12){\line(1,2){02.5}}

\put(45,05){\makebox(0,0)[bl]{(f) cluster \(X_{6}\) }}

\put(80.5,14.5){\oval(09,08)}

\put(78,17){\circle*{1.6}} \put(78,17){\circle{2.6}}
\put(83,17){\circle*{1.0}} \put(83,17){\circle*{1.8}}
\put(78,12){\circle*{1.0}} \put(78,12){\circle*{1.0}}
\put(83,12){\circle*{1.0}} \put(83,12){\circle{1.8}}

\put(78,12){\line(1,0){05}} \put(78,17){\line(1,0){05}}
\put(78,12){\line(0,1){05}} \put(83,12){\line(0,1){05}}

\put(78,12){\line(1,1){05}} \put(83,12){\line(-1,1){05}}

\put(70,05){\makebox(0,0)[bl]{(g) cluster \(X_{7}\) }}

\put(105.5,14.5){\oval(09,08)}

\put(103,17){\circle*{1.0}} \put(103,17){\circle*{1.8}}
\put(108,17){\circle*{1.0}} \put(108,17){\circle{1.8}}
\put(105.5,12){\circle*{1.0}} \put(105.5,12){\circle{1.8}}

\put(103,17){\line(1,0){05}}

\put(105.5,12){\line(-1,2){02.5}} \put(105.5,12){\line(1,2){02.5}}

\put(95,05){\makebox(0,0)[bl]{(h) cluster \(X_{8}\) }}

\end{picture}
\end{center}

 Some illustrative examples  of the problem  solutions are shown in figures:

 (a) clusters balanced by cluster size (about 4),
 tree-like structure over clusters (Fig. 3);

 (b) clusters balanced by  cluster size (about 4)
 and types of elements (three elements of type 1, one element of type 2),
 tree-like structure over clusters (Fig. 4);

 (c) clusters are divided by layers,
 clusters are balanced at each layer (by element types, by element number),
 layered structure over clusters (Fig. 5);
 basic cluster structure by element types:
 bottom layer: three elements of type 4, one element of type 3;
 medium layer: three elements of type 3, one element of type 2;
 top layer: three elements of type 2, one element of type 1.

 Evidently, the quality of the obtained solution
 involves two parts:

 (i) quality of balanced clustering, e.g.,
 proximity for each cluster
 to the specified requirements to the cluster
 (e.g., by element number, by total element weight, by element
 types)
 (e.g., \cite{lev17bal1,lev17balj});

 (ii) quality of the structure over the balanced clusters
 (e.g., proximity to a specified structure
 as chain, tree, balanced tree, hierarchy)
 (e.g., \cite{lev15}).

\begin{center}
\begin{picture}(75,34)
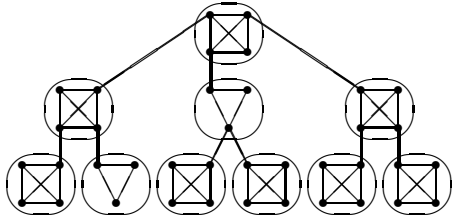
\put(00,00){\makebox(0,0)[bl]{Fig. 3. Example (element numbers
 balance) }}

\put(30.5,29.5){\oval(09,08)}

\put(28,32){\circle*{1.0}} \put(33,32){\circle*{1.0}}
\put(28,27){\circle*{1.0}}\put(33,27){\circle*{1.0}}

\put(28,27){\line(1,0){05}} \put(28,32){\line(1,0){05}}
\put(28,27){\line(0,1){05}} \put(33,27){\line(0,1){05}}

\put(28,27){\line(1,1){05}} \put(33,27){\line(-1,1){05}}

\put(28,27){\line(0,-1){05}} \put(28.2,27){\line(0,-1){05}}
\put(28,32){\line(-3,-2){15}} \put(28,32.2){\line(-3,-2){15}}
\put(33,32){\line(3,-2){15}} \put(33,32.2){\line(3,-2){15}}

\put(10.5,19.5){\oval(09,08)}

\put(08,22){\circle*{1.0}} \put(13,22){\circle*{1.0}}
\put(08,17){\circle*{1.0}} \put(13,17){\circle*{1.0}}

\put(08,17){\line(1,0){05}} \put(08,22){\line(1,0){05}}
\put(08,17){\line(0,1){05}} \put(13,17){\line(0,1){05}}

\put(08,17){\line(1,1){05}} \put(13,17){\line(-1,1){05}}

\put(08,17){\line(0,-1){05}} \put(08.2,17){\line(0,-1){05}}
\put(13,17){\line(0,-1){05}} \put(13.2,17){\line(0,-1){05}}

\put(30.5,19.5){\oval(09,08)}

\put(28,22){\circle*{1.0}} \put(33,22){\circle*{1.0}}
\put(30.5,17){\circle*{1.0}}

\put(30.5,17){\line(-1,2){02.5}} \put(30.5,17){\line(1,2){02.5}}
\put(28,22){\line(1,0){05}}

\put(30.5,17){\line(-1,-2){02.5}}
\put(30.4,17){\line(-1,-2){02.5}}

\put(30.5,17){\line(1,-2){02.5}} \put(30.6,17){\line(1,-2){02.5}}


\put(50.5,19.5){\oval(09,08)}

\put(48,22){\circle*{1.0}} \put(53,22){\circle*{1.0}}
\put(48,17){\circle*{1.0}} \put(53,17){\circle*{1.0}}

\put(48,17){\line(1,0){05}} \put(48,22){\line(1,0){05}}
\put(48,17){\line(0,1){05}} \put(53,17){\line(0,1){05}}

\put(48,17){\line(1,1){05}} \put(53,17){\line(-1,1){05}}

\put(48,17){\line(0,-1){05}} \put(48.2,17){\line(0,-1){05}}
\put(53,17){\line(0,-1){05}} \put(53.2,17){\line(0,-1){05}}

\put(05.5,9.5){\oval(09,08)}

\put(03,12){\circle*{1.0}} \put(08,12){\circle*{1.0}}
\put(03,07){\circle*{1.0}} \put(08,07){\circle*{1.0}}

\put(03,07){\line(1,0){05}} \put(03,12){\line(1,0){05}}
\put(03,07){\line(0,1){05}} \put(08,07){\line(0,1){05}}

\put(03,07){\line(1,1){05}} \put(08,07){\line(-1,1){05}}

\put(15.5,9.5){\oval(09,08)}

\put(13,12){\circle*{1.0}} \put(18,12){\circle*{1.0}}
\put(15.5,07){\circle*{1.0}}


\put(15.5,07){\line(-1,2){02.5}} \put(15.5,07){\line(1,2){02.5}}
\put(13,12){\line(1,0){05}}


\put(25.5,9.5){\oval(09,08)}

\put(23,12){\circle*{1.0}} \put(28,12){\circle*{1.0}}
\put(23,07){\circle*{1.0}} \put(28,07){\circle*{1.0}}

\put(23,07){\line(1,0){05}} \put(23,12){\line(1,0){05}}
\put(23,07){\line(0,1){05}} \put(28,07){\line(0,1){05}}

\put(23,07){\line(1,1){05}} \put(28,07){\line(-1,1){05}}

\put(35.5,9.5){\oval(09,08)}

\put(33,12){\circle*{1.0}} \put(38,12){\circle*{1.0}}
\put(33,07){\circle*{1.0}} \put(38,07){\circle*{1.0}}

\put(33,07){\line(1,0){05}} \put(33,12){\line(1,0){05}}
\put(33,07){\line(0,1){05}} \put(38,07){\line(0,1){05}}

\put(33,07){\line(1,1){05}} \put(38,07){\line(-1,1){05}}

\put(45.5,9.5){\oval(09,08)}

\put(43,12){\circle*{1.0}} \put(48,12){\circle*{1.0}}
\put(43,07){\circle*{1.0}} \put(48,07){\circle*{1.0}}

\put(43,07){\line(1,0){05}} \put(43,12){\line(1,0){05}}
\put(43,07){\line(0,1){05}} \put(48,07){\line(0,1){05}}

\put(43,07){\line(1,1){05}} \put(48,07){\line(-1,1){05}}

\put(55.5,9.5){\oval(09,08)}

\put(53,12){\circle*{1.0}} \put(58,12){\circle*{1.0}}
\put(53,07){\circle*{1.0}} \put(58,07){\circle*{1.0}}

\put(53,07){\line(1,0){05}} \put(53,12){\line(1,0){05}}
\put(53,07){\line(0,1){05}} \put(58,07){\line(0,1){05}}

\put(53,07){\line(1,1){05}} \put(58,07){\line(-1,1){05}}

\end{picture}
%
\begin{picture}(60,34)
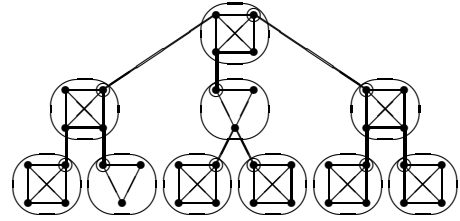
\put(00,00){\makebox(0,0)[bl]{Fig. 4. Example
 (element type balance) }}

\put(30.5,29.5){\oval(09,08)}

\put(28,32){\circle*{1.0}}

\put(33,32){\circle*{1.0}} \put(33,32){\circle{1.8}}

\put(28,27){\circle*{1.0}}\put(33,27){\circle*{1.0}}

\put(28,27){\line(1,0){05}} \put(28,32){\line(1,0){05}}
\put(28,27){\line(0,1){05}} \put(33,27){\line(0,1){05}}

\put(28,27){\line(1,1){05}} \put(33,27){\line(-1,1){05}}

\put(28,27){\line(0,-1){05}} \put(28.2,27){\line(0,-1){05}}
\put(28,32){\line(-3,-2){15}} \put(28,32.2){\line(-3,-2){15}}
\put(33,32){\line(3,-2){15}} \put(33,32.2){\line(3,-2){15}}

\put(10.5,19.5){\oval(09,08)}

\put(08,22){\circle*{1.0}}

\put(13,22){\circle*{1.0}}\put(13,22){\circle{1.8}}

\put(08,17){\circle*{1.0}} \put(13,17){\circle*{1.0}}

\put(08,17){\line(1,0){05}} \put(08,22){\line(1,0){05}}
\put(08,17){\line(0,1){05}} \put(13,17){\line(0,1){05}}

\put(08,17){\line(1,1){05}} \put(13,17){\line(-1,1){05}}

\put(08,17){\line(0,-1){05}} \put(08.2,17){\line(0,-1){05}}
\put(13,17){\line(0,-1){05}} \put(13.2,17){\line(0,-1){05}}

\put(30.5,19.5){\oval(09,08)}

\put(28,22){\circle*{1.0}} \put(28,22){\circle{1.8}}

\put(33,22){\circle*{1.0}} \put(30.5,17){\circle*{1.0}}

\put(30.5,17){\line(-1,2){02.5}} \put(30.5,17){\line(1,2){02.5}}
\put(28,22){\line(1,0){05}}

\put(30.5,17){\line(-1,-2){02.5}}
\put(30.4,17){\line(-1,-2){02.5}}

\put(30.5,17){\line(1,-2){02.5}} \put(30.6,17){\line(1,-2){02.5}}


\put(50.5,19.5){\oval(09,08)}

\put(48,22){\circle*{1.0}} \put(48,22){\circle{1.8}}

\put(53,22){\circle*{1.0}} \put(48,17){\circle*{1.0}}
\put(53,17){\circle*{1.0}}

\put(48,17){\line(1,0){05}} \put(48,22){\line(1,0){05}}
\put(48,17){\line(0,1){05}} \put(53,17){\line(0,1){05}}

\put(48,17){\line(1,1){05}} \put(53,17){\line(-1,1){05}}

\put(48,17){\line(0,-1){05}} \put(48.2,17){\line(0,-1){05}}
\put(53,17){\line(0,-1){05}} \put(53.2,17){\line(0,-1){05}}

\put(05.5,9.5){\oval(09,08)}

\put(03,12){\circle*{1.0}}

\put(08,12){\circle*{1.0}} \put(08,12){\circle{1.8}}

\put(03,07){\circle*{1.0}} \put(08,07){\circle*{1.0}}

\put(03,07){\line(1,0){05}} \put(03,12){\line(1,0){05}}
\put(03,07){\line(0,1){05}} \put(08,07){\line(0,1){05}}

\put(03,07){\line(1,1){05}} \put(08,07){\line(-1,1){05}}

\put(15.5,9.5){\oval(09,08)}

\put(13,12){\circle*{1.0}} \put(13,12){\circle{1.8}}

\put(18,12){\circle*{1.0}} \put(15.5,07){\circle*{1.0}}


\put(15.5,07){\line(-1,2){02.5}} \put(15.5,07){\line(1,2){02.5}}
\put(13,12){\line(1,0){05}}


\put(25.5,9.5){\oval(09,08)}

\put(23,12){\circle*{1.0}}

\put(28,12){\circle*{1.0}} \put(28,12){\circle{1.8}}

\put(23,07){\circle*{1.0}} \put(28,07){\circle*{1.0}}

\put(23,07){\line(1,0){05}} \put(23,12){\line(1,0){05}}
\put(23,07){\line(0,1){05}} \put(28,07){\line(0,1){05}}

\put(23,07){\line(1,1){05}} \put(28,07){\line(-1,1){05}}

\put(35.5,9.5){\oval(09,08)}

\put(33,12){\circle*{1.0}} \put(33,12){\circle{1.8}}

\put(38,12){\circle*{1.0}} \put(33,07){\circle*{1.0}}
\put(38,07){\circle*{1.0}}

\put(33,07){\line(1,0){05}} \put(33,12){\line(1,0){05}}
\put(33,07){\line(0,1){05}} \put(38,07){\line(0,1){05}}

\put(33,07){\line(1,1){05}} \put(38,07){\line(-1,1){05}}

\put(45.5,9.5){\oval(09,08)}

\put(43,12){\circle*{1.0}}

\put(48,12){\circle*{1.0}}\put(48,12){\circle{1.8}}

\put(43,07){\circle*{1.0}} \put(48,07){\circle*{1.0}}

\put(43,07){\line(1,0){05}} \put(43,12){\line(1,0){05}}
\put(43,07){\line(0,1){05}} \put(48,07){\line(0,1){05}}

\put(43,07){\line(1,1){05}} \put(48,07){\line(-1,1){05}}

\put(55.5,9.5){\oval(09,08)}

\put(53,12){\circle*{1.0}}\put(53,12){\circle{1.8}}

\put(58,12){\circle*{1.0}} \put(53,07){\circle*{1.0}}
\put(58,07){\circle*{1.0}}

\put(53,07){\line(1,0){05}} \put(53,12){\line(1,0){05}}
\put(53,07){\line(0,1){05}} \put(58,07){\line(0,1){05}}

\put(53,07){\line(1,1){05}} \put(58,07){\line(-1,1){05}}

\end{picture}
\end{center}

\begin{center}
\begin{picture}(60,36)
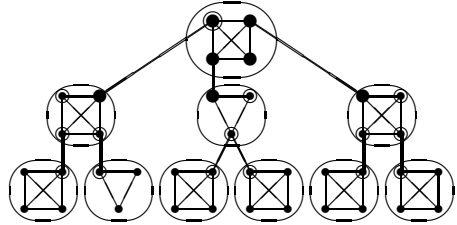
\put(05,00){\makebox(0,0)[bl]{Fig. 5. Three-layer structure }}
\put(30.5,29.5){\oval(12,10)}

\put(28,32){\circle*{1.6}} \put(28,32){\circle{2.6}}
\put(33,32){\circle*{1.0}} \put(33,32){\circle*{1.9}}
\put(28,27){\circle*{1.9}} \put(33,27){\circle*{1.9}}

\put(28,27){\line(1,0){05}} \put(28,32){\line(1,0){05}}
\put(28,27){\line(0,1){05}} \put(33,27){\line(0,1){05}}
\put(28,27){\line(1,1){05}} \put(33,27){\line(-1,1){05}}

\put(28,27){\line(0,-1){05}} \put(28.2,27){\line(0,-1){05}}
\put(28,32){\line(-3,-2){15}} \put(28,32.2){\line(-3,-2){15}}
\put(33,32){\line(3,-2){15}} \put(33,32.2){\line(3,-2){15}}

\put(10.5,19.5){\oval(09,08)}

\put(08,22){\circle*{1.0}} \put(08,22){\circle{1.8}}
\put(13,22){\circle*{1.0}}\put(13,22){\circle*{1.9}}
\put(08,17){\circle*{1.0}} \put(08,17){\circle{1.8}}
\put(13,17){\circle*{1.0}} \put(13,17){\circle{1.8}}
\put(08,17){\line(1,0){05}} \put(08,22){\line(1,0){05}}
\put(08,17){\line(0,1){05}} \put(13,17){\line(0,1){05}}
\put(08,17){\line(1,1){05}} \put(13,17){\line(-1,1){05}}

\put(08,17){\line(0,-1){05}} \put(08.2,17){\line(0,-1){05}}
\put(13,17){\line(0,-1){05}} \put(13.2,17){\line(0,-1){05}}

\put(30.5,19.5){\oval(09,08)}

\put(28,22){\circle*{1.5}} \put(28,22){\circle*{1.9}}
\put(33,22){\circle*{1.0}} \put(33,22){\circle{1.8}}
\put(30.5,17){\circle*{1.0}} \put(30.5,17){\circle{1.8}}
\put(30.5,17){\line(-1,2){02.5}} \put(30.5,17){\line(1,2){02.5}}
\put(28,22){\line(1,0){05}}

\put(30.5,17){\line(-1,-2){02.5}}
\put(30.4,17){\line(-1,-2){02.5}}

\put(30.5,17){\line(1,-2){02.5}} \put(30.6,17){\line(1,-2){02.5}}


\put(50.5,19.5){\oval(09,08)}

\put(48,22){\circle*{1.0}} \put(48,22){\circle*{1.9}}
\put(53,22){\circle*{1.0}} \put(53,22){\circle{1.8}}
\put(48,17){\circle*{1.0}} \put(48,17){\circle{1.8}}
\put(53,17){\circle*{1.0}} \put(53,17){\circle{1.8}}
\put(48,17){\line(1,0){05}} \put(48,22){\line(1,0){05}}
\put(48,17){\line(0,1){05}} \put(53,17){\line(0,1){05}}

\put(48,17){\line(1,1){05}} \put(53,17){\line(-1,1){05}}

\put(48,17){\line(0,-1){05}} \put(48.2,17){\line(0,-1){05}}
\put(53,17){\line(0,-1){05}} \put(53.2,17){\line(0,-1){05}}

\put(05.5,9.5){\oval(09,08)}

\put(03,12){\circle*{1.0}}

\put(08,12){\circle*{1.0}} \put(08,12){\circle{1.8}}
\put(03,07){\circle*{1.0}} \put(08,07){\circle*{1.0}}
\put(03,07){\line(1,0){05}} \put(03,12){\line(1,0){05}}
\put(03,07){\line(0,1){05}} \put(08,07){\line(0,1){05}}

\put(03,07){\line(1,1){05}} \put(08,07){\line(-1,1){05}}

\put(15.5,9.5){\oval(09,08)}

\put(13,12){\circle*{1.0}} \put(13,12){\circle{1.8}}
\put(18,12){\circle*{1.0}} \put(15.5,07){\circle*{1.0}}


\put(15.5,07){\line(-1,2){02.5}} \put(15.5,07){\line(1,2){02.5}}
\put(13,12){\line(1,0){05}}


\put(25.5,9.5){\oval(09,08)}

\put(23,12){\circle*{1.0}}

\put(28,12){\circle*{1.0}} \put(28,12){\circle{1.8}}
\put(23,07){\circle*{1.0}} \put(28,07){\circle*{1.0}}
\put(23,07){\line(1,0){05}} \put(23,12){\line(1,0){05}}
\put(23,07){\line(0,1){05}} \put(28,07){\line(0,1){05}}

\put(23,07){\line(1,1){05}} \put(28,07){\line(-1,1){05}}

\put(35.5,9.5){\oval(09,08)}

\put(33,12){\circle*{1.0}} \put(33,12){\circle{1.8}}
\put(38,12){\circle*{1.0}} \put(33,07){\circle*{1.0}}
\put(38,07){\circle*{1.0}}

\put(33,07){\line(1,0){05}} \put(33,12){\line(1,0){05}}
\put(33,07){\line(0,1){05}} \put(38,07){\line(0,1){05}}
\put(33,07){\line(1,1){05}} \put(38,07){\line(-1,1){05}}

\put(45.5,9.5){\oval(09,08)}

\put(43,12){\circle*{1.0}}

\put(48,12){\circle*{1.0}}\put(48,12){\circle{1.8}}
\put(43,07){\circle*{1.0}} \put(48,07){\circle*{1.0}}

\put(43,07){\line(1,0){05}} \put(43,12){\line(1,0){05}}
\put(43,07){\line(0,1){05}} \put(48,07){\line(0,1){05}}
\put(43,07){\line(1,1){05}} \put(48,07){\line(-1,1){05}}

\put(55.5,9.5){\oval(09,08)}

\put(53,12){\circle*{1.0}}\put(53,12){\circle{1.8}}
\put(58,12){\circle*{1.0}} \put(53,07){\circle*{1.0}}
\put(58,07){\circle*{1.0}}

\put(53,07){\line(1,0){05}} \put(53,12){\line(1,0){05}}
\put(53,07){\line(0,1){05}} \put(58,07){\line(0,1){05}}
\put(53,07){\line(1,1){05}} \put(58,07){\line(-1,1){05}}

\end{picture}
\end{center}

 Consider the evaluation of balanced clusters.
 Let
 \(e^{0} = ( \eta^{0}_{1},...,\eta^{0}_{\xi},...,\eta^{0}_{l} ) \)
 be an estimate of the required structure (by element types) for the balanced
 cluster.
 Thus, proximity between the required structure and cluster
 \(X_{j}\) is (e.g., difference by components):
  \( \delta (e(X_{j}), e^{0})\).
 For simplification, the following definition can be used:~~
%
%
%
 \(\delta (e(X_{j}),e^{0})=\sum_{\xi =\overline{1,l}}~ |\eta_{\xi}
 (X_{j})-\eta^{0}_{\xi}| \).
 For the balance clustering solution \(\widetilde{X}\),
 quality estimate of the cluster balance can be
 considered as follows (for example):~~
 \(Q^{cb} =  \max_{j=\overline{1,k}}  ~~\delta (e(X_{j}),e^{0})
 \).

 The following basic typical spanning structures
 can be used
 \cite{aho83,ander99,chent96,cormen90,gar79,knu97,lev12hier,lev15,stor01}:
%
 (i) various minimum spanning trees (i.e., by total tree weight,
 \(T^{m}\))
 \cite{cher76,cormen90,gabow86,gar79,knu97},
%
 (ii) degree balanced trees (\(T^{d}\),  Fig. 6)
 \cite{cam83,chent96,ran18}
 (including \(k\)-ary trees
 \cite{stor01},
 degree-bounded trees \cite{kone05}),
%
 (iii) height  balanced trees (\(T^{h}\))
 \cite{adel62,aho83,chou09,cormen90,gar79,karl76,knu97}
 (Fig. 7),
%
 (iv) weight balanced trees
 (\(T^{w}\))
 (e.g., balance of the sizes of subtrees in each node)
 \cite{bl80,hir11,niev72},
 and
 (v) special multi-layer structures \cite{lev12hier,lev15}.

\begin{center}
\begin{picture}(70,23)
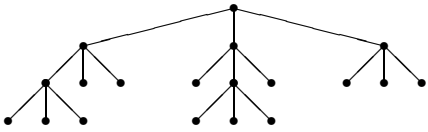
\put(0.5,00){\makebox(0,0)[bl]{Fig. 6. Balanced by degree (3)
 tree }}


\put(10,15){\circle*{1.0}}

\put(05,10){\circle*{1.0}} \put(10,10){\circle*{1.0}}
\put(15,10){\circle*{1.0}}

\put(05,10){\line(1,1){05}} \put(10,10){\line(0,1){05}}
\put(15,10){\line(-1,1){05}}

\put(05,10){\circle*{1.0}}

\put(00,05){\circle*{1.0}} \put(05,05){\circle*{1.0}}
\put(10,05){\circle*{1.0}}

\put(00,05){\line(1,1){05}} \put(05,05){\line(0,1){05}}
\put(10,05){\line(-1,1){05}}

\put(30,15){\circle*{1.0}}

\put(25,10){\circle*{1.0}} \put(30,10){\circle*{1.0}}
\put(35,10){\circle*{1.0}}

\put(25,10){\line(1,1){05}} \put(30,10){\line(0,1){05}}
\put(35,10){\line(-1,1){05}}

\put(30,10){\circle*{1.0}}

\put(25,05){\circle*{1.0}} \put(30,05){\circle*{1.0}}
\put(35,05){\circle*{1.0}}

\put(25,05){\line(1,1){05}} \put(30,05){\line(0,1){05}}
\put(35,05){\line(-1,1){05}}

\put(50,15){\circle*{1.0}}

\put(45,10){\circle*{1.0}} \put(50,10){\circle*{1.0}}
\put(55,10){\circle*{1.0}}

\put(45,10){\line(1,1){05}} \put(50,10){\line(0,1){05}}
\put(55,10){\line(-1,1){05}}

\put(30,20){\circle*{1.0}}

\put(30,20){\line(4,-1){20}} \put(30,20){\line(0,-1){05}}
\put(30,20){\line(-4,-1){20}}

\end{picture}
%
\begin{picture}(80,26)
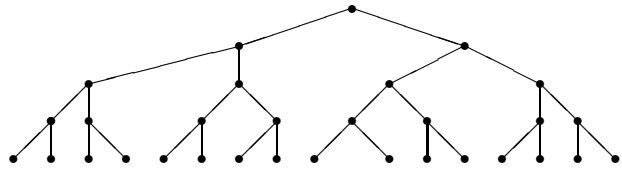
\put(10,00){\makebox(0,0)[bl]{Fig. 7. Balanced by height (3)
 tree }}

\put(10,15){\circle*{1.0}} \put(05,10){\circle*{1.0}}


\put(05,10){\line(1,1){05}}


\put(05,10){\circle*{1.0}}

\put(00,05){\circle*{1.0}} \put(05,05){\circle*{1.0}}
\put(00,05){\line(1,1){05}} \put(05,05){\line(0,1){05}}

\put(10,05){\circle*{1.0}} \put(10,05){\line(0,1){05}}
\put(10,10){\circle*{1.0}} \put(10,10){\line(0,1){05}}

\put(15,05){\circle*{1.0}} \put(15,05){\line(-1,1){05}}

\put(30,15){\circle*{1.0}} \put(25,10){\circle*{1.0}}


\put(35,10){\circle*{1.0}} \put(25,10){\line(1,1){05}}


\put(35,10){\line(-1,1){05}}

\put(20,05){\circle*{1.0}} \put(25,05){\circle*{1.0}}
\put(20,05){\line(1,1){05}}\put(25,05){\line(0,1){05}}

\put(30,05){\circle*{1.0}} \put(35,05){\circle*{1.0}}
\put(30,05){\line(1,1){05}} \put(35,05){\line(0,1){05}}

\put(50,15){\circle*{1.0}} \put(45,10){\circle*{1.0}}


\put(55,10){\circle*{1.0}}

\put(45,10){\line(1,1){05}}


\put(55,10){\line(-1,1){05}}

\put(55,05){\circle*{1.0}} \put(55,05){\line(0,1){05}}

\put(40,05){\circle*{1.0}} \put(50,05){\circle*{1.0}}
\put(40,05){\line(1,1){05}} \put(50,05){\line(-1,1){05}}
\put(60,05){\circle*{1.0}} \put(60,05){\line(-1,1){05}}


\put(65,05){\circle*{1.0}} \put(65,05){\line(1,1){05}}
\put(70,05){\circle*{1.0}}  \put(70,05){\line(0,1){05}}
\put(75,05){\circle*{1.0}} \put(75,05){\line(0,1){05}}
\put(80,05){\circle*{1.0}} \put(80,05){\line(-1,1){05}}

\put(70,10){\circle*{1.0}}  \put(70,10){\line(0,1){05}}
\put(75,10){\circle*{1.0}} \put(75,10){\line(-1,1){05}}

\put(70,15){\circle*{1.0}}  \put(70,15){\line(-2,1){10}}
\put(60,20){\circle*{1.0}} \put(60,20){\line(-2,-1){10}}

\put(30,20){\circle*{1.0}}

\put(30,20){\line(0,-1){05}} \put(30,20){\line(-4,-1){20}}

\put(45,25){\circle*{1.0}}

\put(45,25){\line(-3,-1){15}} \put(45,25){\line(3,-1){15}}

\end{picture}
\end{center}

 Let \(T^{r}\) be a required structure (e.g., the above-mentioned
 type of tree),
 \(T(\widetilde{X})\) be a spanning structure (tree)
 over obtained balanced clusters
 for clustering solution \(\widetilde{X}\).
 Note some proximities of structures (e.g., trees)
 have been briefly described in
 \cite{lev12hier,lev15}.
 Quality estimate of the spanning structure is:~
  \(Q^{s} =   \Delta ( T( \widetilde{X}), T^{r})\).
 As a result,
 the total quality estimate of clustering solution \(\widetilde{X}\)
 with corresponding spanning structure (tree)
  \(T(\widetilde{X})\) is a vector:~~
%
%
%
 \((Q^{cb}, Q^{s})= (\max_{j=\overline{1,k}} \delta
 (e(X_{j}),e^{0}), \Delta (T( \widetilde{X}  ), T^{r}) \).
 Finally,
 the problem of balanced clustering with a required
 spanning structure can be examined as follows:

~~

 Find the balanced clustering solution \(\widetilde{X}\) and corresponding
  spanning structure (tree) for the obtained clusters
 \(T ( \widetilde{X} ) \) such that~
 \( (Q^{cb}, Q^{s}) \Longrightarrow  \min \).

~~

 Evidently, Pareto efficient solutions have to be examined here.

 In the case of multi-layer structures,
 the following evident general solving scheme
 can be considered:
 (1) dividing the initial item set into subsets
 (each subset corresponds to
  solution layer),
 (2) solving the problem above for each layer,
 (3) connection of the layers.
 In some cases, item dividing and spanning stages can be
 integrated (e.g., in maximum leaf spanning tree problem).

 In general, versions of the above-mentioned problem
 are very complicated (its subproblems belong to class of NP-hard
 problems).
 Only some simplifying versions may be  polynomial solvable.
 Thus,
 composite solving schemes (frameworks, strategies) can be used:
 combinations of subproblems as balanced clustering
 and spanning problem(s).
 Note phases of the above-mentioned solving strategies can be based
 on the following:
 (i) heuristics
 (e.g., greedy algorithms, approximation algorithms),
 (ii) enumerative methods,
 (iii) polynomial algorithms
 (for simple cases, e.g., minimum spanning tree).

\section{Basic solving strategies}

 Generally, four basic greedy-like solving strategies can be considered
 (Table 1.).
 Further, the schemes are described. Here
 the cluster balance requirement
 is mainly based on cluster element structure.

\begin{center}
\begin{small}
 {\bf Table 1.} Basic solving strategies\\
%
 \begin{tabular}{| c | l | l |}
\hline
 No.&Strategy &Stages\\

\hline
 1.&Balancing-spanning strategy:
  &(1) balanced clustering,\\
 & &(2) spanning structure (e.g., spanning tree) over clusters\\

 2.&Spanning-balancing strategy:
  &(1) spanning structure (e.g., spanning tree) over items,\\
  &&(2) balanced partitioning the spanning structure  \\
  && (e.g., tree partitioning) \\

 3.&Direct solving strategy:&
   item clustering while taking into account requirements\\
  &&  of cluster balances and spanning structure \\

 4.&Design of balanced layered structure:
   &(1) dividing the item set into subsubsets corresponding
   \\
 && to layers (selection of items/nodes for  each layer),\\

  &&(2) balanced clustering of items at each layer while taking \\
 && into account the requirement of spanning
 structure\\

 &&over the clusters,\\

  && (3) connection of layers
    (including assignment/selection\\

  &&   of the cluster heads for the connection/allocation)\\

\hline
\end{tabular}
\end{small}
\end{center}

\subsection{Balancing-spanning strategy}

 This solving strategy (strategy 1) involves two phases:

 {\it Phase 1.} Balanced clustering of the initial set of elements
   \cite{lev17bal1,lev17balj}.

 {\it Phase 2.} Design of a structure over the obtained balanced clusters
 (e.g., solving some spanning tree problem over the obtained clusters,
 basic  spanning tree problems are pointed out in Table 2).

 The approach above is a basic one and is used in many network
 applications (e.g., design, management, routing).

\begin{center}
\begin{small}
 {\bf Table 2.} Basic spanning problems/methods  \\
 \begin{tabular}{| c | l | l |}
\hline
 No.&Research &Source(s)\\

\hline
 1.&Minimum spanning trees problems: &\\

 1.1.& Basic minimum spanning tree problems
    &\cite{cher76,cormen90,gabow86,gar79,pettie02}\\

 1.2.&Minimum diameter spanning tree problem&\cite{gfe12} \\

 1.3.&Minimum spanning forest problems&\cite{gar79,pettie02a} \\

 1.4.&Minimum spanning multi-tree problems& \cite{geo12,itai89,tar76} \\

 1.5.&Leaf-constrained minimum spanning tree problem&\cite{singh09}\\
   &(similar to maximum leaf spanning tree problem)&\\


\hline
 2.& Multicriteria (multi-objective) spanning tree  problems: &\\

 2.1.&Spanning tree problem with multiple objectives&\cite{hamaher95}\\

 2.2.&Multi-criteria minimum spanning tree problem&\cite{arroyo08,chen07}\\

 2.3.&Combining linear and non-linear objectives in spanning tree problems
    &\cite{dell00}\\

\hline
 3.&Spanning trees with maximum number of leafs problems: &\\

 3.1.&Maximum leaf spanning tree problems
     &\cite{fernau11,fuj03,gar79,lu92,lu98,solis98}\\

 3.2.&Maximum leaf spanning arborescence problem&\cite{drescher10}\\

 3.3.&Spanning trees with many leafs problems&\cite{klei91,solis98}\\

 3.4.& Spanning directed trees with many leaves problem&\cite{alon09}\\

 3.5.& Connected dominating set problems
   (in sense of exact algorithms,
  &\cite{blum05,caro00,cheng03,gar79,thai07}\\

 &  the problem is equivalent
  to maximum leaf spanning tree problem)&\\

\hline
 4.& Balanced spanning tree problems: & \\

 4.1.&Balanced spanning trees problems
  & \cite{aho83,chent96,cormen90,gar79,knu97}\\

 4.2.&Height balanced spanning tree problems & \cite{adel62,aho83,cormen90,gar79,knu97}\\

 4.3.&Degree balanced spanning tree problems&\cite{cam83,chent96,ran18}\\

\hline
\end{tabular}
\end{small}
\end{center}

\newpage
\subsection{Spanning-balancing strategy}

 The strategy (strategy 2) involves two phases:


 {\it Phase 1.} Design of a spanning structure over the initial set of
 elements (e.g., solving some spanning tree problem over the initial set of
 elements) (e.g., Table 2).

 {\it Phase 2.} Design of the balanced clusters
 (e.g., by condensing of the neighbor elements)
 while taking into account the obtained general structure.
 Here, structure-partitioning problem and corresponding methods can be used,
 e.g., tree (hierarchy)-partitioning
 (Table 3).

\begin{center}
\begin{small}
 {\bf Table 3.} Some partitioning of trees/hierarchies approaches \\
 \begin{tabular}{| c | l | l |}
\hline
 No.&Research &Source(s)\\

\hline
 1.&Polynomial algorithms for partitioning a tree into
 single-center subtrees
  &\cite{apol08}\\

  &to minimize flat service costs&\\

 2.&Shifting algorithm techniques for the partitioning of trees
   &\cite{beck95}\\

 3.&Max-min tree partitioning&\cite{perl81}\\

 4.& Uniform centered partitions of trees&\cite{lari16a}\\

 5.&Balanced partitioning of trees&\cite{feld15}\\

 6.&Tree partitioning problem&\cite{mama05}\\

 7.&Graph tree partition problems&\cite{cordo04}\\

 8.&Minimum bounded edge-partition of a tree&\cite{dye09}\\

 9.&Partitioning hierarchically clustered complex networks
  &\cite{meyer16a}\\

 &(via size-constrained graph clustering)&\\

 10.& Optimal hierarchical graph decompositions
   & \cite{racke08}\\

 &(for congestion minimization in network)&\\

 11.&Clustering based on minimum spanning tree&\cite{gry06,lev15c}\\

 12.&Clustering based on minimum and maximum spanning trees &\cite{asa88}\\

 13.&Tree partitions for cryptographic access control&\cite{cram16}\\

\hline
\end{tabular}
\end{small}
\end{center}

 Further,
 four greedy-like solving schemes are described.
 The strategies are based on detection/selection
 of ``condensing''/integration points (node(s) or edge(s) of the examined tree).
 The neighbor node of the ``condensing'' point are integrated with
 the point to get the resultant balanced cluster or its part
 (while taking into account the requirements/rules to the
 cluster).
 The following four types of the ``condensing'' points  are considered:

  (1) minimum weight edge
 (as in agglomerative/hierarchical clustering)
  (Fig. 8);

  (2) leaf nodes: minimum weight edge from a leaf node to  its
  neighbor node
   (bottom-up solving procedure)
   (Fig. 9);

  (3) root node, i.e., minimum weight edge from root node to its neighbor node
  (up-down solving procedure)
   (Fig. 10);

  (4) ``condensing'' points are obtained by using a special location
   procedure (Fig. 11).

 The solving schemes are as follows:

~~

   {\bf Scheme 2.1.} Detection of the minimum weight edge(s)
   (Fig. 8):

 {\it Stage 1.} Analysis of the initial tree.
  Detection of the minimum weight edge \((a',a'')\),
  Usually \(a'\) and \(a''\) are of different types.

 {\it Stage 2.} Building the integrated node: \(J_{a',a''} = a'\&a''\)

 {\it Stage 3.} Addition of the neighbor node to the integrated node
  \(J\) to get the required (by element structure)
  cluster (or quasi-cluster).

 {\it Stage 4.} Correction of the initial tree by deletion of the
  obtained cluster.

 {\it Stage 5.} Analysis of the new tree.
  Go To Stage 1 to examine the obtained tree if it exists,
  otherwise Go To Stage 6.

 {\it Stage 6.} Analysis of the resultant clustering solution.
  In the case of separated nodes, joining
 (addition)
  of the node(s) to the closest
  cluster (or general allocation of the separate nodes to clusters).
  Evaluation of the solution.

 {\it Stage 7.} Stopping.

~~

\newpage

   {\bf Scheme 2.2.} Detection of the minimum weight edge(s)
   for leaf node(s) (Fig. 9):


 {\it Stage 1.}
 Analysis of the initial tree.
 Detection of the leaf node \(b\) with minimum weight edge
 to its neighbor node \(a\);
   it is preferred that  \(b\) and \(a\) are of different types.

 {\it Stage 2.} Building the integrated node: \(J_{b,a} = b\&a\)

 {\it Stage 3.} Addition of the neighbor node to the integrated node
  \(J\) to get the required (by element structure)
  cluster (or quasi-cluster).

 {\it Stage 4.} Correction of the initial tree by deletion of the
  obtained cluster.

 {\it Stage 5.} Analysis of the new tree.
  Go To Stage 1 to examine the obtained tree if it exists,
  otherwise Go To Stage 6.

 {\it Stage 6.} Analysis of the resultant clustering solution.
  In the case of separated nodes, joining (addition)
  of the node(s) to the closest cluster
  (or general allocation of the separate nodes to clusters).
  Evaluation of the solution.

 {\it Stage 7.} Stopping.

~~

  {\bf Scheme 2.3.} Detection of the minimum weight edge(s)
   for root node(s) (Fig. 10):

 {\it Stage 1.}
 Analysis of the initial tree.
 Detection of the neighbor node \(a\) for root \(r\) with minimum weight edge
 between them;
   it is preferred that  \(r\) and \(a\) are of different types.

  {\it Stage 2.} Building the integrated node: \(J_{r,a} = r\&a\).

  {\it Stage 3.} Addition of the neighbor node to the integrated node
  \(J\) to get the required (by element structure)
  cluster (or quasi-cluster).

  {\it Stage 4.} Correction of the initial tree by deletion of the
  obtained cluster.

  {\it Stage 5.} Analysis of the new tree(s).
  Go To Stage 1 to examine the obtained tree if it exists,
  otherwise Go To Stage 6.

 {\it Stage 6.} Analysis of the resultant clustering solution.
  In the case of separated nodes, joining (addition)
  of the node(s) to the closest cluster
  (or general allocation of the separate nodes to clusters).
  Evaluation of the solution.

 {\it Stage 7.} Stopping.

  ~~

  {\bf Scheme 2.4.} Cluster centers based procedure
  (it is similar to \(c\)-mean clustering approach)
    (Fig. 11):

 {\it Stage 1.}
 Analysis of the initial tree.
 Detection of a set of special nodes
(e.g., nodes of type 1)
 as cluster centers
  \(c^{1},...,c^{\gamma},...,c^{\lambda}\).

 {\it Stage 2.} Detection of cluster center \(c^{\gamma}\)
 that have its neighbor node \(a\)
 (another node type is preferable)
  with the minimum edge weight.

 {\it Stage 3.} Building the integrated node: \(J_{c^{\gamma},a} =c^{\gamma} \&a\)

  {\it Stage 4.} Addition of the neighbor node to the integrated node
  \(J\) to get the required (by element structure)
  cluster (or quasi-cluster).

  {\it Stage 5.} Correction of the initial tree by deletion of the
  obtained cluster.

 {\it Stage 6.} Analysis of the new tree(s).
  Go To Stage 1 to examine the obtained tree if it exists,
  otherwise Go To Stage 7.

 {\it Stage 7.} Analysis of the resultant clustering solution.
  In the case of separated nodes, joining (addition) of the node(s) to the closest
  cluster
  (or general allocation of the separate nodes to clusters).
  Evaluation of the solution.

 {\it Stage 8.} Stopping.

~~

 It is reasonable to point out the following:

 {\it Note 1.} The above-mentioned solving schemes are polynomial
 (their complexities equal \(O(n^{2})\) or less).

  {\it Note 2.} Evidently, the described solving schemes
 can be transform into parallel procedures
 (i.e., parallel examination and usage of the ``condensing'' points, etc.).

\begin{center}
\begin{picture}(75,40)
\put(00,00){\makebox(0,0)[bl]{Fig. 8. Minimum weight edge}}

\put(02,30){\makebox(0,0)[bl]{Basic}}
\put(02,27){\makebox(0,0)[bl]{spanning}}
\put(02,24){\makebox(0,0)[bl]{tree}}
\put(16,28){\vector(2,-1){06}}

\put(00,10){\line(1,0){66}} \put(00,10){\line(3,2){33}}
\put(66,10){\line(-3,2){33}}

\put(23.5,31.3){\makebox(0,0)[bl]{Root}}
\put(33,32){\circle*{0.7}} \put(33,32){\circle{2.2}}
\put(34.5,31.7){\makebox(0,0)[bl]{\(r\)}}

\put(20,18){\circle*{2.0}} 
\put(25,13){\circle*{0.8}} \put(25,13){\circle{1.4}}

\put(10,13){\circle*{0.8}} \put(10,13){\circle{1.4}}
\put(20,18){\line(-2,-1){10}}

\put(17,14){\makebox(0,0)[bl]{\(a'\)}}
\put(20,11.5){\makebox(0,0)[bl]{\(a''\)}}

\put(21.5,15.5){\oval(14,09)}

\put(20,18){\line(1,-1){05}}

\put(30,17){\makebox(0,0)[bl]{Minimum}}
\put(30,14){\makebox(0,0)[bl]{weight}}
\put(30,11){\makebox(0,0)[bl]{edge}}

\put(29.5,16){\vector(-1,0){06.7}}

\end{picture}
%
\begin{picture}(70,40)
\put(00,00){\makebox(0,0)[bl]{Fig. 9. Leaf node with
 minimum weight edge}}

\put(02,30){\makebox(0,0)[bl]{Basic}}
\put(02,27){\makebox(0,0)[bl]{spanning}}
\put(02,24){\makebox(0,0)[bl]{tree}}
\put(16,28){\vector(2,-1){06}}

\put(00,10){\line(1,0){66}} \put(00,10){\line(3,2){33}}
\put(66,10){\line(-3,2){33}}

\put(23.5,31.3){\makebox(0,0)[bl]{Root}}
\put(33,32){\circle*{0.7}}\put(33,32){\circle{2.2}}
\put(34.5,31.7){\makebox(0,0)[bl]{\(r\)}}

\put(20,15){\circle*{0.8}} \put(20,15){\circle{1.4}}
\put(25,10){\circle*{1.0}} 

\put(10,10){\circle*{1.0}} 
\put(20,15){\line(-2,-1){10}}

\put(16.5,14.5){\makebox(0,0)[bl]{\(a\)}}
\put(21,10.5){\makebox(0,0)[bl]{\(b\)}}

\put(21.5,13){\oval(14,08)}

\put(20,15){\line(1,-1){05}}

\put(30,17){\makebox(0,0)[bl]{Minimum}}
\put(30,14){\makebox(0,0)[bl]{weight}}
\put(30,11){\makebox(0,0)[bl]{edge}}

\put(29.5,15){\vector(-2,-1){05.7}}

\end{picture}
\end{center}

\begin{center}
\begin{picture}(80,40)
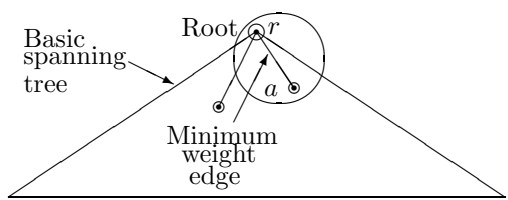
\put(00,00){\makebox(0,0)[bl]{Fig. 10. Root with
 minimum weight edge}}

\put(02,30){\makebox(0,0)[bl]{Basic}}
\put(02,27){\makebox(0,0)[bl]{spanning}}
\put(02,24){\makebox(0,0)[bl]{tree}}
\put(16,28){\vector(2,-1){06}}

\put(00,10){\line(1,0){66}} \put(00,10){\line(3,2){33}}
\put(66,10){\line(-3,2){33}}

\put(23,31.5){\makebox(0,0)[bl]{Root}}
\put(33,32){\circle*{0.7}}\put(33,32){\circle{2.2}}

\put(34.5,31.7){\makebox(0,0)[bl]{\(r\)}}

\put(38,24.5){\circle*{0.8}} \put(38,24.5){\circle{1.4}}
\put(33,32){\line(2,-3){05}} 
\put(34,23.5){\makebox(0,0)[bl]{\(a\)}}

\put(28,22){\circle*{0.8}} \put(28,22){\circle{1.4}}

\put(33,32){\line(-1,-2){05}}

\put(36,28.5){\oval(12,12)}

\put(21,17){\makebox(0,0)[bl]{Minimum}}
\put(23,14){\makebox(0,0)[bl]{weight}}
\put(24,11){\makebox(0,0)[bl]{edge}}

\put(30,20){\vector(1,2){04.5}}

\end{picture}
%
\begin{picture}(75,40)
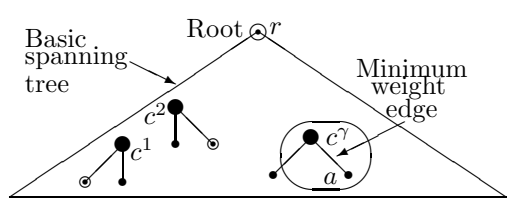
\put(00,00){\makebox(0,0)[bl]{Fig. 11. Cluster center, minimum
 weight edge}}

\put(02,30){\makebox(0,0)[bl]{Basic}}
\put(02,27){\makebox(0,0)[bl]{spanning}}
\put(02,24){\makebox(0,0)[bl]{tree}}
\put(16,28){\vector(2,-1){06}}

\put(00,10){\line(1,0){66}} \put(00,10){\line(3,2){33}}
\put(66,10){\line(-3,2){33}}

\put(23.5,31.3){\makebox(0,0)[bl]{Root}}
\put(33,32){\circle*{0.7}} \put(33,32){\circle{2.2}}
\put(34.5,31.7){\makebox(0,0)[bl]{\(r\)}}

\put(15,17){\circle*{2.0}} 
\put(15,12){\circle*{1.0}} 

\put(10,12){\circle*{0.8}} \put(10,12){\circle{1.4}}
\put(15,17){\line(-1,-1){05}}

\put(16,15){\makebox(0,0)[bl]{\(c^{1}\)}}
\put(15,17){\line(0,-1){05}}

\put(22,22){\circle*{2.0}}

\put(27,17){\circle*{0.8}} \put(27,17){\circle{1.4}}

\put(22,17){\circle*{1.0}} 
\put(22,22){\line(0,-1){05}}

\put(17.8,19.7){\makebox(0,0)[bl]{\(c^{2}\)}}
\put(22,22){\line(1,-1){05}}

\put(40,18){\circle*{2.0}} 
\put(45,13){\circle*{1.0}} 

\put(35,13){\circle*{1.0}} 
\put(40,18){\line(-1,-1){05}} 

\put(42,16.9){\makebox(0,0)[bl]{\(c^{\gamma}\)}}
\put(41.7,11.8){\makebox(0,0)[bl]{\(a\)}}

\put(42,15.5){\oval(12,09)}

\put(40,18){\line(1,-1){05}}

\put(46,26){\makebox(0,0)[bl]{Minimum}}
\put(48,23){\makebox(0,0)[bl]{weight}}
\put(50,20){\makebox(0,0)[bl]{edge}}
\put(52.5,20){\vector(-2,-1){09}}

\end{picture}
\end{center}

 Now an illustrative numerical example is considered:

 (a) an initial element (item) set and their parameters are contained in
 Table 4:
  \(A = \{a_{1},...,a_{i},...,a_{21} \}\);

 (b) the edge weights between item pairs
 \(a_{i_{1}}\) and \(a_{i_{2}}\)
 (\( 1 \leq i_{1} < \leq i_{2}  \leq 18\))
 are contained in Table 5;

 (c) a basic (obtained) minimum spanning tree structure
 (by edge weights/proximities) is shown in Fig. 12.

\begin{center}
\begin{small}
 {\bf Table 4.} Elements and their parameters  \\
 \begin{tabular}{| c | c | c| c|}
\hline
 No.&Item &Item type &Number of balanced cluster  \\

   &\(a_{i}\)&\(\xi\)& \(X_{j}\) (in solution \(\widetilde{X}\))  \\

\hline

 1.&\(a_{1}\) & \(1\) &\(6\)  \\

 2.&\(a_{2}\) & \(2\) &\(6\)  \\

 3.&\(a_{3}\) & \(1\) &\(6\)  \\

 4.&\(a_{4}\) & \(1\) &\(2\)  \\

 5.&\(a_{5}\) & \(1\) &\(5\)  \\

 6.&\(a_{6}\) & \(1\) &\(3\)  \\

 7.&\(a_{7}\) & \(2\) &\(2\)  \\

 8.&\(a_{8}\) & \(3\) &\(2\)  \\

 9.&\(a_{9}\) & \(2\) &\(5\)  \\

 10.&\(a_{10}\) & \(3\) &\(5\)  \\

 11.&\(a_{11}\) & \(2\) &\(3\)  \\

 12.&\(a_{12}\) & \(1\) &\(3\)  \\

 13.&\(a_{13}\) & \(1\) &\(4\)  \\

 14.&\(a_{14}\) & \(1\) &\(1\)  \\

 15.&\(a_{15}\) & \(2\) &\(4\)  \\

 16.&\(a_{16}\) & \(3\) &\(4\)  \\

 17.&\(a_{17}\) & \(2\) &\(1\)  \\

 18.&\(a_{18}\) & \(3\) &\(1\)  \\

 19.&\(a_{19}\) & \(3\) &\(1\)  \\



\hline
\end{tabular}
\end{small}
\end{center}

\newpage
\begin{center}
\begin{small}
{\bf Table 5.} Edge weights between
  items  \(a_{i_{1}}\) and   \(a_{i_{2}}\)    \\
\begin{tabular}{| c |cc c c c c c c c c c c c c ccccc|}
\hline
 \(i_{1}\)  & \(i_{2}:\)
  & \(2\)&\(3\)&\(4\)&\(5\)&\(6\) &\(7\)&\(8\)&\(9\)&\(10\)
  &\(11\)&\(12\) &\(13\)&\(14\)&\(15\)&\(16\) &\(17\)&\(18\)
  &\(19\)  \\

\hline

 1 &&\(2.5\)&\(2.8\)&\(\star \)&\(\star \)&\(\star\)&\(\star\)&\(\star\)&\(\star\)&\(\star\)
  &\(\star\)&\(\star\)&\(\star\)&\(\star\)
  &\(\star\)&\(\star\)&\(\star\)&\(\star\) &\(\star\) \\

 2 &&&\(\star\)&\(3.0\)&\(\star\)&\(\star\)&\(\star\)&\(\star\)
 &\(\star\)&\(\star\)&\(\star\)&\(\star\)&\(\star\)&\(\star\)
  &\(\star\)&\(\star\)&\(\star\)&\(\star\)&\(\star\) \\

 3 &&&&\(\star\)&\(3.5\)&\(4.1\)&\(\star\)&\(\star\)&\(\star\)
  &\(\star\)&\(\star\)&\(\star\)&\(\star\)& \(\star\)
  &\(\star\)&\(\star\)&\(\star\)&\(\star\) &\(\star\)\\

 4 &&&&&\(\star\)&\(\star\)&\(1.0\)&\(0.6\)&\(\star\)
 &\(\star\)&\(\star\)&\(\star\)&\(\star\)& \(\star\)
 &\(\star\)&\(\star\)&\(\star\)&\(\star\)&\(\star\) \\

 5 &&&&&&\(\star\)&\(\star\)&\(\star\)&\(1.3\)&\(1.2\)
  &\(\star\)&\(\star\)&\(\star\)& \(\star\)
 &\(\star\)&\(\star\)&\(\star\)&\(\star\)&\(\star\) \\

 6 &&&&&&&\(\star\)&\(\star\)&\(\star\)&\(\star\)
 &\(1.1\)&\(1.0\)&\(\star\)& \(\star\)
 &\(\star\)&\(\star\)&\(\star\)&\(\star\)&\(\star\) \\

 7 &&&&&&&&\(\star\)&\(\star\)&\(\star\)&\(\star\)&\(\star\)&\(4.2\)& \(\star\)
  &\(\star\)&\(\star\)&\(\star\)&\(\star\)&\(\star\) \\

 8 &&&&&&&&&\(\star\) &\(\star\)&\(\star\)&\(\star\)
 & \(\star\)&\(\star\)
  &\(\star\)&\(\star\)&\(\star\)&\(\star\)&\(\star\) \\

 9 &&&&&&&&&&\(\star\)&\(\star\)&\(\star\)&\(\star\)&\(\star\)
 &\(\star\)&\(\star\)&\(\star\)&\(\star\)&\(\star\) \\

 10 &&&&&&&&&&&\(\star\)&\(\star\)&\(\star\)&\(\star\)
 &\(\star\)&\(\star\)&\(\star\)&\(\star\) &\(\star\)\\

 11 &&&&&&&&&&&&\(\star\)&\(\star\)&\(\star\)
 &\(\star\)&\(\star\)&\(\star\)&\(\star\)&\(\star\) \\

 12 &&&&&&&&&&&&&\(\star\)&\(\star\)
 &\(\star\)&\(\star\)&\(\star\)&\(\star\)&\(\star\) \\

 13 &&&&&&&&&&&&&&\(\star\)
 &\(1.1\)&\(1.3\)&\(\star\)&\(\star\)&\(\star\) \\

 14 &&&&&&&&&&&&&&&\(\star\)
  &\(\star\)&\(1.0\)&\(0.5\)&\(2.0\)\\

 15 &&&&&&&&&&&&&&&&\(\star\)
  &\(\star\)&\(\star\) &\(\star\)\\

 16 &&&&&&&&&&&&&&&&&\(\star\)
 &\(\star\)&\(\star\) \\

 17 &&&&&&&&&&&&&&&&&&\(\star\)&\(\star\)\\

 18 &&&&&&&&&&&&&&&&&&&\(\star\)\\

\hline
\end{tabular}
\end{small}
\end{center}

\begin{center}
\begin{picture}(77,59)
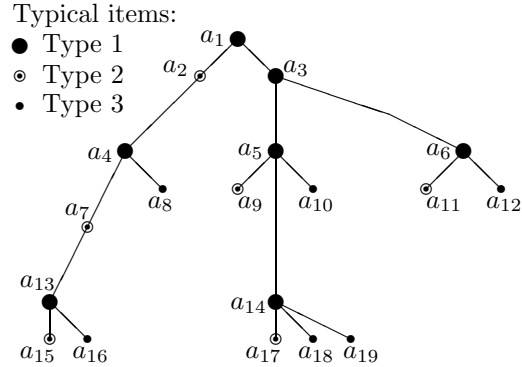
\put(12,00){\makebox(0,0)[bl]{Fig. 12. Basic spanning tree}}

\put(00,51.5){\makebox(0,0)[bl]{Typical items: }}

\put(01,49){\circle*{2.0}}

\put(04,47.5){\makebox(0,0)[bl]{Type \(1\) }}

\put(01,45){\circle*{0.8}} \put(01,45){\circle{1.4}}
\put(04,43.5){\makebox(0,0)[bl]{Type \(2\) }}

\put(01,41){\circle*{1.0}}

\put(04,39.5){\makebox(0,0)[bl]{Type \(3\) }}


\put(05,15){\circle*{2.0}}
\put(01,16.5){\makebox(0,0)[bl]{\(a_{13}\)}}

\put(05,10){\circle*{0.8}} \put(05,10){\circle{1.4}}
\put(05,10){\line(0,1){05}}
\put(01,07){\makebox(0,0)[bl]{\(a_{15}\)}}

\put(10,10){\circle*{1.0}} \put(10,10){\line(-1,1){05}}
\put(08,07){\makebox(0,0)[bl]{\(a_{16}\)}}


\put(15,35){\circle*{2.0}}
\put(10,33.5){\makebox(0,0)[bl]{\(a_{4}\)}}

\put(10,25){\circle*{0.8}} \put(10,25){\circle{1.4}}
\put(10,25){\line(1,2){05}}
\put(07,26){\makebox(0,0)[bl]{\(a_{7}\)}}

\put(20,30){\circle*{1.0}} \put(20,30){\line(-1,1){05}}
\put(018,27){\makebox(0,0)[bl]{\(a_{8}\)}}

\put(10,25){\line(-1,-2){05}}


\put(35,15){\circle*{2.0}}
\put(29,13.5){\makebox(0,0)[bl]{\(a_{14}\)}}

\put(35,10){\circle*{0.8}} \put(35,10){\circle{1.4}}
\put(35,10){\line(0,1){05}}
\put(31,07){\makebox(0,0)[bl]{\(a_{17}\)}}

\put(40,10){\circle*{1.0}} \put(40,10){\line(-1,1){05}}
\put(38,07){\makebox(0,0)[bl]{\(a_{18}\)}}

\put(45,10){\circle*{1.0}} \put(45,10){\line(-2,1){10}}
\put(44,07){\makebox(0,0)[bl]{\(a_{19}\)}}


\put(35,35){\circle*{2.0}}
\put(30,34){\makebox(0,0)[bl]{\(a_{5}\)}}

\put(30,30){\circle*{0.8}} \put(30,30){\circle{1.4}}
\put(30,30){\line(1,1){05}}
\put(30,27){\makebox(0,0)[bl]{\(a_{9}\)}}

\put(40,30){\circle*{1.0}} \put(40,30){\line(-1,1){05}}
\put(38,27){\makebox(0,0)[bl]{\(a_{10}\)}}

\put(35,35){\line(0,-1){20}}


\put(30,50){\circle*{2.0}}
\put(25,49){\makebox(0,0)[bl]{\(a_{1}\)}}

\put(25,45){\circle*{0.8}} \put(25,45){\circle{1.4}}
\put(25,45){\line(1,1){05}}
\put(20,45){\makebox(0,0)[bl]{\(a_{2}\)}}

\put(35,45){\circle*{2.0}} \put(35,45){\line(-1,1){05}}
\put(36,45){\makebox(0,0)[bl]{\(a_{3}\)}}

\put(35,45){\line(0,-1){15}}

\put(35,45){\line(3,-1){15}} \put(50,40){\line(2,-1){10}}

\put(25,45){\line(-1,-1){10}}


\put(60,35){\circle*{2.0}}
\put(55,34){\makebox(0,0)[bl]{\(a_{6}\)}}

\put(55,30){\circle*{0.8}} \put(55,30){\circle{1.4}}
\put(55,30){\line(1,1){05}}
\put(55,27){\makebox(0,0)[bl]{\(a_{11}\)}}

\put(65,30){\circle*{1.0}} \put(65,30){\line(-1,1){05}}
\put(63,27){\makebox(0,0)[bl]{\(a_{12}\)}}



\end{picture}
\end{center}

 It is assumed
  the required structure of the balanced cluster is:
   node of type 1, node of type 2, node of type 3.
   The corresponding estimate of the cluster structure
   equals ~ \( e^{0} = (1,1,1)  \).
 The basic set of leaf nodes is:~~
 \(L = \{a_{8},a_{9},a_{10},a_{11},a_{12},a_{15},a_{16},a_{17},a_{18},a_{19}\} \)
 (Table 6).

\begin{center}
\begin{small}
{\bf Table 6.} Leaf nodes and corresponding distances\\
%
\begin{tabular}{| c |c |c|}
\hline
 Leaf node \(a_{i}\) & Edge \((a_{i},b)\) & Edge weight  \(w(a_{i},b)\) \\

\hline

 I. Basic leaf nodes:&&\\

 \(a_{8}\) & \((a_{8},a_{4}) \) &\(0.6\)\\

 \(a_{9}\) & \((a_{9},a_{5}) \) &\(1.3\)\\

 \(a_{10}\) & \((a_{10},a_{5}) \) &\(1.2\)\\

 \(a_{11}\) & \((a_{11},a_{6}) \) &\(1.3\)\\

 \(a_{12}\) & \((a_{12},a_{6}) \) &\(1.0\)\\

 \(a_{15}\) & \((a_{15},a_{13}) \) &\(1.1\)\\

 \(a_{16}\) & \((a_{16},a_{13}) \) &\(1.3\)\\

 \(a_{17}\) & \((a_{17},a_{14}) \) &\(1.0\)\\

 \(a_{18}\) & \((a_{18},a_{14}) \) &\(0.5\)\\

 \(a_{19}\) & \((a_{19},a_{14}) \) &\(2.0\)\\

\hline

 II. Additional leaf nodes:&&\\

 \(a_{2}\) & \((a_{2},a_{1}) \) &\(2.5\)\\

 \(a_{3}\) & \((a_{3},a_{1}) \) &\(2.8\)\\

\hline
\end{tabular}
\end{small}
\end{center}

\newpage
 Solving scheme 2.2 is used as follows:

~~

  {\it 1st algorithm step:}~ Selection of the minimum weight edge:
  \((a_{18}, a_{14}) \)
  (weight \(w(a_{i},b)=0.5\)).
  Integration of nodes:  \(J_{a_{18},a_{14}} = a_{18} \& a_{14}\)
  (an initial part of cluster).
 Evident addition of node \(a_{17}\) to the integrated node to
 obtain a balanced (by element structure)
 cluster \(X_{1} = \{ a_{14},a_{17},a_{18} \}\),
 estimate of the cluster structure is:
  \(e(X_{1}) = (1,1,1)\).

  {\it 2nd algorithm step:}~ Selection of the minimum weight edge:
  \((a_{8}, a_{4}) \)
  (weight \(w(a_{i},b)=0.6\)).
  Integration of nodes:  \(J_{a_{8},a_{4}} = a_{8} \& a_{4}\)
  (an initial part of cluster).
 Evident addition of node \(a_{7}\) to the integrated node to
 obtain a balanced (by element structure)
 cluster \(X_{2} = \{ a_{4},a_{7},a_{8} \}\),
 estimate of the cluster structure is:
  \(e(X_{2}) = (1,1,1)\).

 Note,  \(a_{2}\) is an additional leaf node
 (addition in Table 24).

   {\it 3rd algorithm step:}~ Selection of the minimum weight edge:
  \((a_{12}, a_{6}) \)
  (weight \(w(a_{i},b)=1.0\)).
  Integration of nodes:  \(J_{a_{12},a_{6}} = a_{12} \& a_{6}\)
  (an initial part of cluster).
 Evident addition of node \(a_{11}\) to the integrated node to
 obtain a balanced (by element structure)
 cluster \(X_{3} = \{ a_{6},a_{11},a_{12} \}\),
 estimate of the cluster structure is:
  \(e(X_{3}) = (1,1,1)\).

 {\it 4th algorithm step:}~ Selection of the minimum weight edge:
  \((a_{15}, a_{13}) \)
  (weight \(w(a_{i},b)=1.1\)).
  Integration of nodes:  \(J_{a_{15},a_{13}} = a_{15} \& a_{13}\)
  (an initial part of cluster).
 Evident addition of node \(a_{16}\) to the integrated node to
 obtain a balanced (by element structure)
 cluster \(X_{4} = \{ a_{13},a_{15},a_{16} \}\),
 estimate of the cluster structure is:
  \(e(X_{2}) = (1,1,1)\).

 {\it 5th algorithm step:}~ Selection of the minimum weight edge:
  \((a_{10}, a_{5}) \)
  (weight \(w(a_{i},b)=1.2\)).
  Integration of nodes:  \(J_{a_{10},a_{5}} = a_{10} \& a_{5}\)
  (an initial part of cluster).
 Evident addition of node \(a_{9}\) to the integrated node to
 obtain a balanced (by element structure)
 cluster \(X_{5} = \{ a_{5},a_{9},a_{10} \}\),
 estimate of the cluster structure is:
  \(e(X_{5}) = (1,1,1)\).

 Note,  \(a_{3}\) is an additional leaf node
 (addition in Table 24).
 The obtained tree is shown in Fig. 13.

  {\it 6th algorithm step:}~
 Selection of the minimum weight edge:
  \((a_{2}, a_{1}) \)
  (weight \(w(a_{i},b)=2.5\)).
  Integration of nodes:  \(J_{a_{2},a_{1}} = a_{2} \& a_{1}\)
  (an initial part of cluster).
 Evident addition of node \(a_{3}\) to the integrated node to
 obtain a balanced (by element structure)
 cluster \(X_{6} = \{ a_{1},a_{1},a_{3} \}\),
 estimate of the cluster structure is:
  \(e(X_{5}) = (2,1,0)\).

 Note, the obtained cluster has not the required element structure.

 {\it 7th algorithm step:}~
  Extension of cluster \(X_{1}\)
  by the separated node \(a_{19}\):~
 \(\overline{ X}_{1} =  X_{1}  \& a_{19} \),
 \(e(\overline{X}_{1}) = (1,1,2)\).

~~

 Thus, the obtained clustering solution is:~~~
 \(\widetilde{X}=\{\overline{X}_{1},X_{2},X_{3},X_{4},X_{5},X_{6}\} \)
 (Fig. 14, Table 4).

\begin{center}
\begin{picture}(80,54)
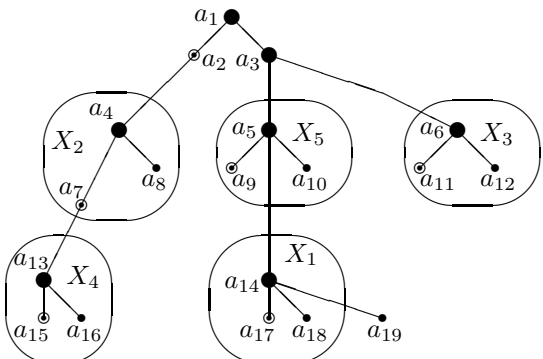
\put(00,00){\makebox(0,0)[bl]{Fig. 13. Spanning tree
 (after algorithm step 5)}}

\put(08,14){\makebox(0,0)[bl]{\(X_{4}\)}}
\put(07,12.5){\oval(14,17)}

\put(05,15){\circle*{2.0}}
\put(01,16.5){\makebox(0,0)[bl]{\(a_{13}\)}}

\put(05,10){\circle*{0.8}} \put(05,10){\circle{1.4}}
\put(05,10){\line(0,1){05}}
\put(01,07){\makebox(0,0)[bl]{\(a_{15}\)}}

\put(10,10){\circle*{1.0}} \put(10,10){\line(-1,1){05}}
\put(08,07){\makebox(0,0)[bl]{\(a_{16}\)}}

\put(06,32){\makebox(0,0)[bl]{\(X_{2}\)}}
\put(014,31.5){\oval(18,17)}

\put(15,35){\circle*{2.0}}
\put(011,36.5){\makebox(0,0)[bl]{\(a_{4}\)}}

\put(10,25){\circle*{0.8}} \put(10,25){\circle{1.4}}
\put(10,25){\line(1,2){05}}
\put(07,26){\makebox(0,0)[bl]{\(a_{7}\)}}

\put(20,30){\circle*{1.0}} \put(20,30){\line(-1,1){05}}
\put(018,27){\makebox(0,0)[bl]{\(a_{8}\)}}

\put(10,25){\line(-1,-2){05}}

\put(37,17){\makebox(0,0)[bl]{\(X_{1}\)}}
\put(36,12.5){\oval(18,17)}

\put(35,15){\circle*{2.0}}
\put(29,13.5){\makebox(0,0)[bl]{\(a_{14}\)}}

\put(35,10){\circle*{0.8}} \put(35,10){\circle{1.4}}
\put(35,10){\line(0,1){05}}
\put(31,07){\makebox(0,0)[bl]{\(a_{17}\)}}

\put(40,10){\circle*{1.0}} \put(40,10){\line(-1,1){05}}
\put(38,07){\makebox(0,0)[bl]{\(a_{18}\)}}

\put(50,10){\circle*{1.0}} \put(50,10){\line(-3,1){15}}
\put(48,07){\makebox(0,0)[bl]{\(a_{19}\)}}

\put(38,33){\makebox(0,0)[bl]{\(X_{5}\)}}
\put(37,32){\oval(18,14)}

\put(35,35){\circle*{2.0}}
\put(30,34){\makebox(0,0)[bl]{\(a_{5}\)}}

\put(30,30){\circle*{0.8}} \put(30,30){\circle{1.4}}
\put(30,30){\line(1,1){05}}
\put(30,27){\makebox(0,0)[bl]{\(a_{9}\)}}

\put(40,30){\circle*{1.0}} \put(40,30){\line(-1,1){05}}
\put(38,27){\makebox(0,0)[bl]{\(a_{10}\)}}

\put(35,35){\line(0,-1){20}}


\put(30,50){\circle*{2.0}}
\put(25,49){\makebox(0,0)[bl]{\(a_{1}\)}}

\put(25,45){\circle*{0.8}} \put(25,45){\circle{1.4}}
\put(25,45){\line(1,1){05}}
\put(26,43){\makebox(0,0)[bl]{\(a_{2}\)}}

\put(35,45){\circle*{2.0}} \put(35,45){\line(-1,1){05}}
\put(30.5,43){\makebox(0,0)[bl]{\(a_{3}\)}}

\put(35,45){\line(0,-1){15}}

\put(35,45){\line(3,-1){15}} \put(50,40){\line(2,-1){10}}

\put(25,45){\line(-1,-1){10}}

\put(63,33){\makebox(0,0)[bl]{\(X_{3}\)}}
\put(62,32){\oval(18,14)}

\put(60,35){\circle*{2.0}}
\put(55,34){\makebox(0,0)[bl]{\(a_{6}\)}}

\put(55,30){\circle*{0.8}} \put(55,30){\circle{1.4}}
\put(55,30){\line(1,1){05}}
\put(55,27){\makebox(0,0)[bl]{\(a_{11}\)}}

\put(65,30){\circle*{1.0}} \put(65,30){\line(-1,1){05}}
\put(63,27){\makebox(0,0)[bl]{\(a_{12}\)}}


\end{picture}
%
\begin{picture}(74,54)
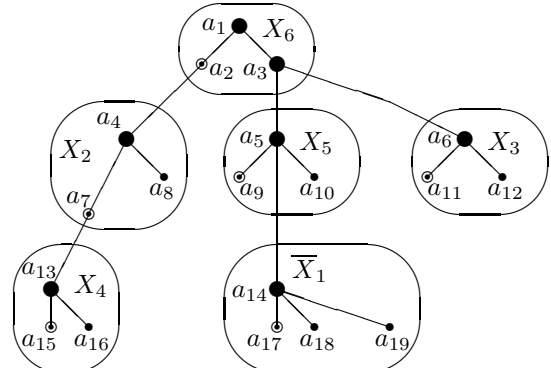
\put(00,00){\makebox(0,0)[bl]{Fig. 14. Spanning tree
 (after algorithm step 7)}}

\put(08,14){\makebox(0,0)[bl]{\(X_{4}\)}}
\put(07,12.5){\oval(14,17)}

\put(05,15){\circle*{2.0}}
\put(01,16.5){\makebox(0,0)[bl]{\(a_{13}\)}}

\put(05,10){\circle*{0.8}} \put(05,10){\circle{1.4}}
\put(05,10){\line(0,1){05}}
\put(01,07){\makebox(0,0)[bl]{\(a_{15}\)}}

\put(10,10){\circle*{1.0}} \put(10,10){\line(-1,1){05}}
\put(08,07){\makebox(0,0)[bl]{\(a_{16}\)}}

\put(06,32){\makebox(0,0)[bl]{\(X_{2}\)}}
\put(014,31.5){\oval(18,17)}

\put(15,35){\circle*{2.0}}
\put(011,36.5){\makebox(0,0)[bl]{\(a_{4}\)}}

\put(10,25){\circle*{0.8}} \put(10,25){\circle{1.4}}
\put(10,25){\line(1,2){05}}
\put(07,26){\makebox(0,0)[bl]{\(a_{7}\)}}

\put(20,30){\circle*{1.0}} \put(20,30){\line(-1,1){05}}
\put(018,27){\makebox(0,0)[bl]{\(a_{8}\)}}

\put(10,25){\line(-1,-2){05}}

\put(37,16){\makebox(0,0)[bl]{\(\overline{X}_{1}\)}}
\put(41,12.5){\oval(26,17)}

\put(35,15){\circle*{2.0}}
\put(29,13.5){\makebox(0,0)[bl]{\(a_{14}\)}}

\put(35,10){\circle*{0.8}} \put(35,10){\circle{1.4}}
\put(35,10){\line(0,1){05}}
\put(31,07){\makebox(0,0)[bl]{\(a_{17}\)}}

\put(40,10){\circle*{1.0}} \put(40,10){\line(-1,1){05}}
\put(38,07){\makebox(0,0)[bl]{\(a_{18}\)}}

\put(50,10){\circle*{1.0}} \put(50,10){\line(-3,1){15}}
\put(48,07){\makebox(0,0)[bl]{\(a_{19}\)}}

\put(38,33){\makebox(0,0)[bl]{\(X_{5}\)}}
\put(37,32){\oval(18,14)}

\put(35,35){\circle*{2.0}}
\put(30,34){\makebox(0,0)[bl]{\(a_{5}\)}}

\put(30,30){\circle*{0.8}} \put(30,30){\circle{1.4}}
\put(30,30){\line(1,1){05}}
\put(30,27){\makebox(0,0)[bl]{\(a_{9}\)}}

\put(40,30){\circle*{1.0}} \put(40,30){\line(-1,1){05}}
\put(38,27){\makebox(0,0)[bl]{\(a_{10}\)}}

\put(35,35){\line(0,-1){20}}

\put(33,48){\makebox(0,0)[bl]{\(X_{6}\)}}
\put(31,47){\oval(18,12)}

\put(30,50){\circle*{2.0}}
\put(25,49){\makebox(0,0)[bl]{\(a_{1}\)}}

\put(25,45){\circle*{0.8}} \put(25,45){\circle{1.4}}
\put(25,45){\line(1,1){05}}
\put(26,43){\makebox(0,0)[bl]{\(a_{2}\)}}

\put(35,45){\circle*{2.0}} \put(35,45){\line(-1,1){05}}
\put(30.5,43){\makebox(0,0)[bl]{\(a_{3}\)}}

\put(35,45){\line(0,-1){15}}

\put(35,45){\line(3,-1){15}} \put(50,40){\line(2,-1){10}}

\put(25,45){\line(-1,-1){10}}

\put(63,33){\makebox(0,0)[bl]{\(X_{3}\)}}
\put(62,32){\oval(18,14)}

\put(60,35){\circle*{2.0}}
\put(55,34){\makebox(0,0)[bl]{\(a_{6}\)}}

\put(55,30){\circle*{0.8}} \put(55,30){\circle{1.4}}
\put(55,30){\line(1,1){05}}
\put(55,27){\makebox(0,0)[bl]{\(a_{11}\)}}

\put(65,30){\circle*{1.0}} \put(65,30){\line(-1,1){05}}
\put(63,27){\makebox(0,0)[bl]{\(a_{12}\)}}

\end{picture}
\end{center}

 The quality of the balanced clustering can be calculated as
  follows \cite{lev17bal1,lev17balj}
 (it is assumed \( X_{1} = \overline{X}_{1}\)):
 \[Q^{cb} (\widetilde{X}) =  \max_{j=\overline{1,6}}  ~~\delta
 (e(X_{j}),e^{0})=
   \max [\delta (e(X_{1}),e^{0})), \delta (e(X_{6}),e^{0} )] =
   \max  [1,2] = 2. \]

\newpage
\subsection{Direct solving strategy}

 This strategy can be considered as hierarchical
 (agglomerative) clustering (bottom-up procedure)
 while taking into account special balance requirements
 (e.g, by cluster size, by cluster element structure).
 Some hierarchical clustering approaches are
 pointed out in Table 7.

\begin{center}
\begin{small}
 {\bf Table 7.} Some hierarchical clustering studies (surveys, methods)\\
%
 \begin{tabular}{| c | l | l |}
\hline
 No.&Research &Some sources\\

\hline
 1.&General surveys on clustering&\cite{jain88,jain99}\\

 2.&Surveys on combinatorial clustering &\cite{lev15c,lev15d}\\

 3.&Approximate hierarchical clustering via sparsest cut and
 spreading metrics
   &\cite{chari17}\\

 4.& Performance guarantees for hierarchical clustering
   &\cite{das05}\\

 5.&Cost function for similarity-based hierarchical clustering
    &\cite{das16}\\

 6.&General approach for incremental approximation and hierarchical
 clustering
   & \cite{lin10}\\

 7.&Hierarchical clustering based on ordinal estimates
    &\cite{lev07e,lev15c}\\

 8.&Hierarchical clustering algorithm based on the Hungarian method
   &\cite{gold08}\\

  9.&Graph-based hierarchical conceptual clustering&\cite{jon01}\\

 10.&Hierarchical graph decompositions (in network)&\cite{racke08}\\

 11.&Hierarchical agglomerative clustering (in
 wireless sensor networks)
   &\cite{lungc10}\\

 12.&Hierarchical clustering for categorical data
 (using a probabilistic rough set model)
   &\cite{lim14}\\

 13.&Hierarchical clustering via joint between-within distances
   &\cite{szekely05}\\

 &(extending Ward's minimum variance method)&\\

 14.&Interval competitive agglomeration clustering algorithm
  &\cite{jeng10}\\

 15.&Unsupervised multidimensional hierarchical clustering&\cite{dug98}\\

\hline
\end{tabular}
\end{small}
\end{center}

 In  \cite{lev15c},
 the procedure and numerical example
 of hierarchical balanced clustering
 (while taking into account cluster sizes)
  are described.

 Note,
 the procedure
 of hierarchical balanced clustering
 while taking into account cluster element structure
 is similar to the above-mentioned procedure.
 On the other hand,
 this strategy is similar to strategy 2.2 from the previous section.

\subsection{Design of layered structures with balancing}

 Here the following
  evident three phase solving strategy can be considered:

  {\it Phase 1.} Dividing the initial set of elements to obtain
  several subset (each obtained element subset corresponds to a
  specified solution layer).
  Here, various procedures can be used, for example:
  (i) selection of the elements,
  (ii) special spanning problems (e.g., maximum leaf node spanning
  tree problem, independent set problem).

  {\it Phase 2.} Solving a special balanced clustering problem
  at each layer.

  {\it Phase 3.} Connection of the layers
  (i.e., connecting the elements of different layers).

 A illustrative example of the balanced three layered structure
 is depicted in Fig. 15 (four types of items are considered).
 Several solving schemes for the design of
 layered structured (including layered networks)
 are described in
 \cite{lev12hier,lev15}.
 Note, some multi-node cluster
 at top network layers
 can be considered as composite hubs
 (e.g., communication networks, transportation networks).

\begin{center}
\begin{picture}(105,62)
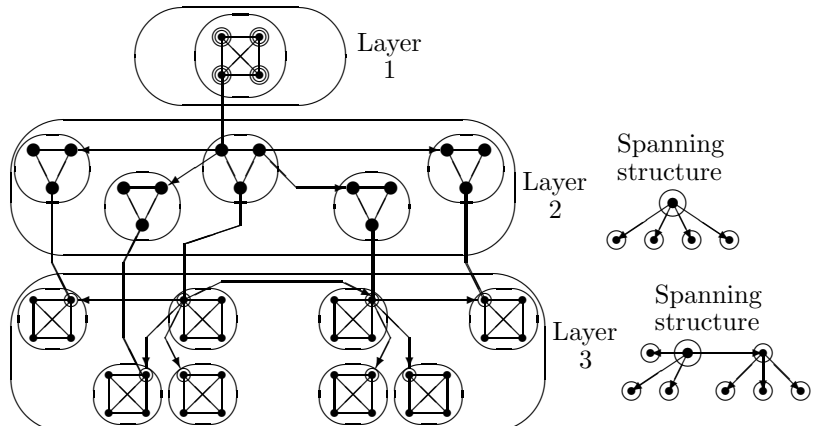
\put(16,00){\makebox(0,0)[bl]{Fig. 15. Example of balanced layered
 structure }}

\put(30.5,54.5){\oval(28,13)}

\put(46,54.5){\makebox(0,0)[bl]{Layer}}
\put(49.5,51.5){\makebox(0,0)[bl]{1}}

\put(30.5,54.5){\oval(12,11)}

\put(28,57){\circle*{0.9}}\put(28,57){\circle{1.8}}
\put(28,57){\circle{2.5}}

\put(33,57){\circle*{0.9}}\put(33,57){\circle{1.8}}
\put(33,57){\circle{2.5}}

\put(28,52){\circle*{0.9}}\put(28,52){\circle{1.8}}
\put(28,52){\circle{2.5}}

\put(33,52){\circle*{0.9}}\put(33,52){\circle{1.8}}
\put(33,52){\circle{2.5}}

\put(28,52){\line(1,0){05}} \put(28,57){\line(1,0){05}}
\put(28,52){\line(0,1){05}} \put(33,52){\line(0,1){05}}

\put(28,52){\line(1,1){05}} \put(33,52){\line(-1,1){05}}

\put(28,52){\line(0,-1){10}} \put(28.1,52){\line(0,-1){10}}

\put(33.5,37){\oval(67,18)}

\put(68,36){\makebox(0,0)[bl]{Layer}}
\put(71.5,33){\makebox(0,0)[bl]{2}}

\put(80.6,41){\makebox(0,0)[bl]{Spanning }}
\put(80.6,38){\makebox(0,0)[bl]{structure}}

\put(88,35){\circle*{1.5}} \put(88,35){\circle{3.5}}

\put(80.5,30){\circle*{1.0}} \put(80.5,30){\circle{2.5}}
\put(88,35){\vector(-3,-2){07.5}}

\put(85.5,30){\circle*{1.0}} \put(85.5,30){\circle{2.5}}
\put(88,35){\vector(-1,-2){02.5}}

\put(95.5,30){\circle*{1.0}} \put(95.5,30){\circle{2.5}}
\put(88,35){\vector(3,-2){07.5}}

\put(90.5,30){\circle*{1.0}} \put(90.5,30){\circle{2.5}}
\put(88,35){\vector(1,-2){02.5}}

\put(05.5,39.5){\oval(10,09)}

\put(03,42){\circle*{1.5}} \put(03,42){\circle*{1.9}}
\put(08,42){\circle*{1.0}} \put(08,42){\circle*{1.9}}
\put(05.5,37){\circle*{1.0}} \put(05.5,37){\circle*{1.9}}

\put(05.5,37){\line(-1,2){02.5}} \put(05.5,37){\line(1,2){02.5}}
\put(03,42){\line(1,0){05}}

\put(05.5,37){\line(0,-1){10}} \put(05.4,37){\line(0,-1){10}}

\put(05.4,27){\line(1,-2){02.5}} \put(05.5,27){\line(1,-2){02.5}}

\put(30.5,37){\line(0,-1){05}} \put(30.6,37){\line(0,-1){05}}

\put(17.5,34.5){\oval(10,09)}

\put(15,37){\circle*{1.5}} \put(15,37){\circle*{1.9}}
\put(20,37){\circle*{1.0}} \put(20,37){\circle*{1.9}}
\put(17.5,32){\circle*{1.0}} \put(17.5,32){\circle*{1.9}}

\put(17.5,32){\line(-1,2){02.5}} \put(17.5,32){\line(1,2){02.5}}
\put(15,37){\line(1,0){05}}

\put(17.5,32){\line(-1,-2){02.5}}
\put(17.4,32){\line(-1,-2){02.5}}

\put(14.9,27){\line(0,-1){10}} \put(15,27){\line(0,-1){10}}

\put(14.9,17){\line(1,-2){02.5}} \put(15,17){\line(1,-2){02.55}}


\put(30.5,39.5){\oval(10,09)}

\put(28,42){\circle*{1.5}} \put(28,42){\circle*{1.9}}
\put(33,42){\circle*{1.0}} \put(33,42){\circle*{1.9}}
\put(30.5,37){\circle*{1.0}} \put(30.5,37){\circle*{1.9}}

\put(30.5,37){\line(-1,2){02.5}} \put(30.5,37){\line(1,2){02.5}}
\put(28,42){\line(1,0){05}}


\put(30.5,37){\line(0,-1){05}} \put(30.6,37){\line(0,-1){05}}

\put(30.5,32){\line(-3,-1){07.5}}
\put(30.6,32){\line(-3,-1){07.5}}

\put(23,29.5){\line(0,-1){07.3}}
\put(23.1,29.5){\line(0,-1){07.3}}

\put(28,42){\vector(-1,0){19}} \put(33,42){\vector(1,0){24}}

\put(28,42){\vector(-3,-2){07}}

\put(33,42){\line(1,-1){05}} \put(38,37){\vector(1,0){06.5}}

\put(48,34.5){\oval(10,09)}

\put(45.5,37){\circle*{1.5}} \put(45.5,37){\circle*{1.9}}
\put(50.5,37){\circle*{1.0}} \put(50.5,37){\circle*{1.9}}
\put(48,32){\circle*{1.0}} \put(48,32){\circle*{1.9}}

\put(48,32){\line(-1,2){02.5}} \put(48,32){\line(1,2){02.5}}
\put(45.5,37){\line(1,0){05}}

\put(48,32){\line(0,-1){10}} \put(48.2,32){\line(0,-1){10}}


\put(60.5,39.5){\oval(10,09)}

\put(58,42){\circle*{1.5}} \put(58,42){\circle*{1.9}}
\put(63,42){\circle*{1.0}} \put(63,42){\circle*{1.9}}
\put(60.5,37){\circle*{1.0}} \put(60.5,37){\circle*{1.9}}

\put(60.5,37){\line(-1,2){02.5}} \put(60.5,37){\line(1,2){02.5}}
\put(58,42){\line(1,0){05}}


\put(60.4,37){\line(0,-1){10}} \put(60.6,37){\line(0,-1){10}}

\put(60.4,27){\line(1,-2){02.5}} \put(60.6,27){\line(1,-2){02.5}}

\put(35.5,15){\oval(71,21)}

\put(72,16){\makebox(0,0)[bl]{Layer}}
\put(75.5,13){\makebox(0,0)[bl]{3}}

\put(85.6,21){\makebox(0,0)[bl]{Spanning }}
\put(85.6,18){\makebox(0,0)[bl]{structure}}

\put(90,15){\circle*{1.5}} \put(90,15){\circle{3.5}}
\put(90,15){\vector(1,0){10}}

\put(85,15){\circle*{1.0}} \put(85,15){\circle{2.5}}
\put(90,15){\vector(-1,0){05}}

\put(82.5,10){\circle*{1.0}} \put(82.5,10){\circle{2.5}}
\put(90,15){\vector(-3,-2){07.5}}

\put(87.5,10){\circle*{1.0}} \put(87.5,10){\circle{2.5}}
\put(90,15){\vector(-1,-2){02.5}}

\put(100,15){\circle*{1.0}} \put(100,15){\circle{2.5}}

\put(95,10){\circle*{1.0}} \put(95,10){\circle{2.5}}
\put(100,15){\vector(-1,-1){05}}

\put(100,10){\circle*{1.0}} \put(100,10){\circle{2.5}}
\put(100,15){\vector(0,-1){05}}

\put(105,10){\circle*{1.0}} \put(105,10){\circle{2.5}}
\put(100,15){\vector(1,-1){05}}

\put(05.5,19.5){\oval(09,08)}

\put(03,22){\circle*{1.0}}

\put(08,22){\circle*{1.0}} \put(08,22){\circle{1.8}}

\put(03,17){\circle*{1.0}} \put(08,17){\circle*{1.0}}

\put(03,17){\line(1,0){05}} \put(03,22){\line(1,0){05}}
\put(03,17){\line(0,1){05}} \put(08,17){\line(0,1){05}}

\put(03,17){\line(1,1){05}} \put(08,17){\line(-1,1){05}}

\put(25.5,19.5){\oval(09,08)}

\put(23,22){\circle*{1.0}} \put(23,22){\circle{1.8}}

\put(28,22){\circle*{1.0}} \put(23,17){\circle*{1.0}}
\put(28,17){\circle*{1.0}}

\put(23,17){\line(1,0){05}} \put(23,22){\line(1,0){05}}
\put(23,17){\line(0,1){05}} \put(28,17){\line(0,1){05}}

\put(23,17){\line(1,1){05}} \put(28,17){\line(-1,1){05}}

\put(23,22){\vector(-1,0){14}}

\put(23,22){\line(-1,-1){05}} \put(18,17){\vector(0,-1){04}}

\put(23,22){\line(-1,-2){02.5}} \put(20.5,17){\vector(1,-2){02}}

\put(23,22){\line(2,1){05}}

\put(28,24.5){\line(1,0){16}} \put(44,24.5){\vector(2,-1){04}}

\put(45.5,19.5){\oval(09,08)}

\put(43,22){\circle*{1.0}}

\put(48,22){\circle*{1.0}}\put(48,22){\circle{1.8}}

\put(43,17){\circle*{1.0}} \put(48,17){\circle*{1.0}}

\put(43,17){\line(1,0){05}} \put(43,22){\line(1,0){05}}
\put(43,17){\line(0,1){05}} \put(48,17){\line(0,1){05}}

\put(43,17){\line(1,1){05}} \put(48,17){\line(-1,1){05}}

\put(48,22){\vector(1,0){14}}

\put(48,22){\line(1,-1){05}} \put(53,17){\vector(0,-1){04}}
\put(48,22){\line(1,-2){02.5}} \put(50.5,17){\vector(-1,-2){02}}

\put(65.5,19.5){\oval(09,08)}

\put(63,22){\circle*{1.0}}\put(63,22){\circle{1.8}}

\put(68,22){\circle*{1.0}} \put(63,17){\circle*{1.0}}
\put(68,17){\circle*{1.0}}

\put(63,17){\line(1,0){05}} \put(63,22){\line(1,0){05}}
\put(63,17){\line(0,1){05}} \put(68,17){\line(0,1){05}}

\put(63,17){\line(1,1){05}} \put(68,17){\line(-1,1){05}}

\put(15.5,09.5){\oval(09,08)}

\put(13,12){\circle*{1.0}}

\put(18,12){\circle*{1.0}} \put(18,12){\circle{1.8}}
\put(13,07){\circle*{1.0}} \put(18,07){\circle*{1.0}}

\put(13,07){\line(1,0){05}} \put(13,12){\line(1,0){05}}
\put(13,07){\line(0,1){05}} \put(18,07){\line(0,1){05}}

\put(13,07){\line(1,1){05}} \put(18,07){\line(-1,1){05}}

\put(25.5,9.5){\oval(09,08)}

\put(23,12){\circle*{1.0}} \put(23,12){\circle{1.8}}

\put(28,12){\circle*{1.0}} \put(23,07){\circle*{1.0}}
\put(28,07){\circle*{1.0}}

\put(23,07){\line(1,0){05}} \put(23,12){\line(1,0){05}}
\put(23,07){\line(0,1){05}} \put(28,07){\line(0,1){05}}

\put(23,07){\line(1,1){05}} \put(28,07){\line(-1,1){05}}

\put(45.5,9.5){\oval(09,08)}

\put(43,12){\circle*{1.0}}

\put(48,12){\circle*{1.0}}\put(48,12){\circle{1.8}}
\put(43,07){\circle*{1.0}} \put(48,07){\circle*{1.0}}

\put(43,07){\line(1,0){05}} \put(43,12){\line(1,0){05}}
\put(43,07){\line(0,1){05}} \put(48,07){\line(0,1){05}}

\put(43,07){\line(1,1){05}} \put(48,07){\line(-1,1){05}}

\put(55.5,9.5){\oval(09,08)}

\put(53,12){\circle*{1.0}}\put(53,12){\circle{1.8}}

\put(58,12){\circle*{1.0}} \put(53,07){\circle*{1.0}}
\put(58,07){\circle*{1.0}}

\put(53,07){\line(1,0){05}} \put(53,12){\line(1,0){05}}
\put(53,07){\line(0,1){05}} \put(58,07){\line(0,1){05}}

\put(53,07){\line(1,1){05}} \put(58,07){\line(-1,1){05}}

\end{picture}
\end{center}

\section{On improvement/correction of solutions}

 The improvement/correction procedure of preliminary obtained solution
 is important additional part of the
 solving strategies.
 In Table 8,
 the basic problem elements
 which are under modification operations in
 improvement/correction approaches
 are pointed out.
 Four basic approaches
 for solution improvement in combinatorial optimization
  are pointed out in  Table 9.

\begin{center}
\begin{small}
 {\bf Table 8.} Basic problem element modification operations\\
\begin{tabular}{| c | l| l |}
\hline
 No.&Element under modification & Operations\\

\hline
 1.&Element as node/vertex (e.g., graph, network)
   &1.Deletion\\

   &&2.Addition\\

   &&3.Condencing (unification)\\

   &&4.Decoupling (disintegration) \\

 2.&Element as edge/arc (e.g., graph, network)
   &1.Deletion\\

   &&2.Addition\\

   &&3.Condencing (unification)\\

   &&4.Decoupling (disintegration) \\

 3.&Element group, cluster
    &1.Deletion\\

   &&2.Addition\\

   &&3.Condencing (unification)\\

   &&4.Decoupling (disintegration) \\

 3.& Structure over elements/clusters
    &1.Augmentation\\

   &(e.g., spanning structure over elements/clusters,
   &2.Decoupling (disintegration)\\

  &i.e., tree, chain, forest, hierarchy, etc.)
  &3.Transformation to another structure\\

\hline
\end{tabular}
\end{small}
\end{center}

\begin{center}
\begin{small}
 {\bf Table 9.} Solution improvement/correction approaches\\
\begin{tabular}{| c | l |l|}
\hline
 No.&Research direction& Source(s)\\

\hline

 1.& Improvement algorithm as part of local optimization and
 heuristic
    &\cite{aiex05,conw67,cormen90,gar79,ho14,moran08,tana94}\\

 &procedures (e.g., exchange techniques, reassignment/recoloring)&\\

 2.&Augmentation problems as modification of an initial graph/network
 &\cite{esw76,fomin17,frank11,gar79,jack05}\\

 & (e.g., addition/correction of solution components as vertices or/and &\\

 &edges/arcs) to get the solution (i.e., graph/network)
 with a required  &\\

 &property(ies), e.g., the required level of  the graph/network connectivity&\\

 4.&Independence free graphs and vertex connectivity augmentation
    & \cite{jack05}\\

 3.& Reoptimization (small correction of an initial solution to improve its
 &\cite{arch13,aus09,bilo08,boria10}\\

  & quality as improvement of the solution objective function(s))&\\

 4.& Restructuring problems (modification of an initial solution while
    &\cite{lev11restr,lev15restr}\\

 &taking into account two criteria: minimum modification cost,   &\\
 &minimum proximity to a next solution at the next time stage)&\\

\hline
\end{tabular}
\end{small}
\end{center}

\section{Conclusion}

 The paper describes the author outline of
 heuristic solving frameworks to the design
 of  balanced clustering
 solutions and  spanning structures (e.g., tree) over the obtained clusters.
 Four basic greedy-like solving strategies are described.
 Examples illustrate the problems and solving procedures.
 Generally, the paper has a preliminary character.
 The material has to be extended in the future.
 It may be reasonable to point out
 some prospective research directions:
 (1)   examination and usage of
  multicriteria and fuzzy problem formulations;
 (2) consideration of other types of spanning structures over the
 obtained balanced clusters  (e.g., chain, forest, hierarchy);
 (3) study of other solving approaches
 (e.g., random algorithms);
 (4) design of a software system for implementation of the
 considered problems and solving schemes;
 (5) evaluation of the solving schemes for the considered problems
  on the basis of computing experiments;
  and
 (6) investigation of various realistic
  network applications of the described clustering
  problems (e.g., network design and management).

\section{Acknowledgments}

 The research was done in
 Institute for Information Transmission Problems of Russian Academy of Sciences
 and supported by the Russian Science Foundation
 (grant 14-50-00150, ``Digital technologies and their
applications'').


\end{document}